\documentclass[reprint,superscriptaddress,floatfix,showpacs,balancelastpage,aps,pra]{revtex4-1}

\usepackage{graphicx}
\usepackage{amsmath}
\usepackage{amssymb}
\usepackage{color}
\usepackage{bbold}

\newcommand{\bra}[1]{\ensuremath{\left\langle #1\right|}}
\newcommand{\ket}[1]{\ensuremath{\left|#1\right\rangle}}
\newcommand{\Cre}[0]{\ensuremath{\hat{a}^{\dagger}}}
\newcommand{\Ann}[0]{\ensuremath{\hat{a}}}
\newcommand{\OP}[1]{\ensuremath{\hat{#1}}}
\newcommand{\Sig}[1]{\ensuremath{\hat\sigma_\mathrm{#1}}}
\newcommand{\Gee}[2]{\ensuremath{g^\mathrm{#1}_{#2}}}

\newcommand{\eff}[1]{\ensuremath{#1_{\mathrm{eff}}}}
\newcommand{\SP}[0]{\ensuremath{\hat\sigma^+}}
\newcommand{\SM}[0]{\ensuremath{\hat\sigma^-}}
\newcommand{\expup}[1]{\ensuremath{\mathrm{e}^{#1}}}

\begin{document}
\title{Multiphoton resonances for all optical quantum logic with multiple cavities}

\author{Mark S.\ Everitt}
\affiliation{National Institute of Informatics, 2-1-2 Hitotsubashi, Chiyoda-ku, Tokyo 101-8430, Japan}
\author{Barry M.\ Garraway}
\affiliation{Department of Physics and Astronomy,
  University of Sussex, Falmer, Brighton, BN1 9QH,
  United Kingdom}

\date{\today.}

\begin{abstract}
  We develop a theory for the interaction of multi-level atoms with
  multi-mode cavities yielding cavity-enhanced multi-photon resonances. The
  locations of the resonances are predicted from the use of effective two-\
  and three-level Hamiltonians.  As an application we show that quantum
  gates can be realised when photonic qubits are encoded on the cavity modes
  in arrangements where ancilla atoms transit the cavity.  The fidelity
  of operations is increased by conditional measurements on the atom and by
  the use of a selected, dual-rail, Hilbert space. A universal set of gates
  is proposed, including the Fredkin gate and i\textsc{swap} operation; the system
  seems promising for scalability.
\end{abstract}

\date{\today}
\maketitle

\section{Introduction}

The field of quantum computation \cite{Nielsen2000} has attracted proposals
for many different physical realisations. Amongst these, the use of photonic
qubits is appealing because of the potential interface to optical
communications, the accessibility of coherent sources for qubits and the
possibility of manipulating those qubits using established optical
technology. With photonic qubits a central issue is that of enabling
sufficiently strong, and coherent, interactions for quantum logic. The field
of cavity QED (or CQED) naturally meets these requirements as it has a
history of coherent quantum interactions and entanglement generation
\cite{Raimond2001,Walther2006}.  Alternatively, non-linear media can be used
for qubit interactions, but very strong non-linearities are needed for the
non-linear media (see e.g.\ Refs.\ \cite{Milburn1989,Yavuz2005,OBrien2007}).
The approach of linear optical computing \cite{Knill2001}, which uses
passive optical components, can be used and seems promising \cite{Kok2007},
although the use of ``flying qubits,'' generally pulses encoded in
polarisation states, can make the approach susceptible to photon losses
\cite{OBrien2007}, and it also places high demands on single-photon sources.

With CQED a single photon can have a strong interaction with an atom, which
usually results in at least a two step process for interactions between
photonic qubits. This approach has been used for quantum logic with, for
example, flying photonic qubits and cavities \cite{Duan2004}, with qubits in
atoms that talk via a cavity mode ``bus'' \cite{Pellizzari1995}, and with
qubits in both the atoms and the cavity modes
\cite{Rauschenbeutel1999,Lin2006,Biswas2004,Joshi2006}.  If we wanted to
avoid the losses associated with flying photonic qubits, and use cavity
storage of photons, we can still use CQED with an atom ``bus''.  Examples
typically involve two cavity modes (with qubits represented as the absence
or presence of a photon) and three-level, or more complex, atoms (see e.g.\
\cite{Zubairy2003,Shu2007,Lin2008}).

Existing methods for quantum logic with stored cavity photons do not, to our
knowledge, use the advantageous ``dual-rail'' channel \cite{Chuang1995,
  Kok2007}.  With flying qubits the dual-rail approach means that a qubit is
typically encoded as a single-photon pulse in one of two polarisations. This
means that qubit loss is detected by the absence of a photon. We will adapt
this approach to cavity stored photons by encoding a single qubit on a pair
of cavity field modes. The presence of excitation in a first mode (no
excitation in the second) encodes the \ket{1}\ qubit state, and the presence
of excitation in the second mode encodes a qubit state \ket{0} (see Table
\ref{Table: qubit}).
\begin{table}[t]
\centering
\begin{tabular}{ccc}
Modes && Qubit\\
\hline
\ket{1}\ket{0} &$\mapsto$& \ket{1}\\
\ket{0}\ket{1} &$\mapsto$& \ket{0}\\
\end{tabular}
\caption{A qubit is encoded as a single excitation shared between two modes of the field. 
   A logical one maps to the excitation residing fully in the first mode, and
   the logical zero maps to the excitation being fully in the second mode. 
   \label{Table: qubit}}
\end{table}
Thus superpositions of qubit states will involve entangled states of the
cavity modes.  In Sec.\ \ref{sec:rots} of this paper we show how to achieve
the $x$-rotation of single logical qubits which can be arranged to ensure
that, for example, $\ket{0}\longrightarrow(\ket{0}+\ket{1})/\sqrt{2}$.
However, in terms of \emph{physical} states, an entangled state
$(\ket{01}+\ket{10})/\sqrt{2}$ has been created. There has been considerable
interest in such states recently in terms of the decay of entanglement
\cite{Bellomo07,Maniscalco08,Wang08,Li09,Mazzola09a,Zhou09,Ferraro09,%
Francica09,Xiao09,Zhang09,Mazzola09c,Mazzola10b,Mazzola10a,%
Fanchini10,Ge10,Wang10,Man10}.  The $x$-rotation procedure would
create such an entangled state with photons in independent reservoirs.
An interesting feature of the qubit encoding (Table~\ref{Table: qubit}) is
that, if any excitation is lost, then the resulting state of the pair of
modes, $\ket{0}\ket{0}$, does not map to a valid qubit. In this way quantum
information processing amounts to the rearrangement of photons in our
system.

\begin{figure}[ht]
\includegraphics[width=0.45\textwidth]{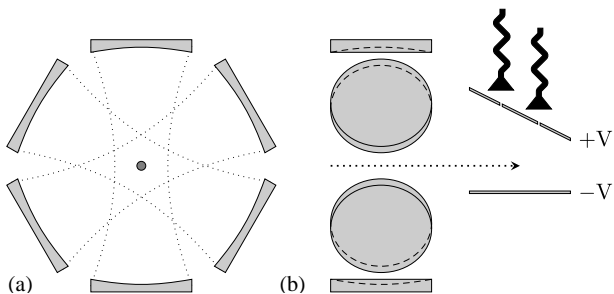}
\caption{Simplified illustration of the physical layout of a single gate of the type
considered in the Paper. Axial (a) and side (b) views are shown.  A single
atom enters a multi-mode cavity and interacts with the photonic qubits
present. An interaction with as many as six cavity modes is considered. On
exit from the cavities, the atomic state is measured. The illustration shows
state selective field ionization \cite{Meschede1985} as an example. The
measurement allows us to enhance the fidelity of the gate operation.
\label{figure1}}
\end{figure}

The clear disadvantage of this approach, in a CQED implementation, is that
the duplexity of cavity modes increases the difficulty of formulating gates,
and increases the vulnerability to cavity decay. The purpose of this paper
is to show that, nevertheless, a practical and \emph{universal} set of gates
can be found for dual-rail qubits in a CQED system using strong coupling. We
will see that the rearrangement of photons is performed by an ancilla atom
which enters and then leaves the multi-mode cavity (see Fig.\
\ref{figure1}).
Because the quantum information resides in the cavity modes, the ancilla
atom must leave in an un-entangled state.  (This is the inverse of the case
where two atoms interact with a detuned cavity field to produce entanglement
between the atoms, while remaining un-entangled with the field
\cite{Zheng2000,Osnaghi2001}.) We can ``help'' the disentanglement by
performing a measurement on the atom when it has left the cavity.  The
measurement is intended to ``clean up'' the quantum state, i.e.\ the
probability of failure is low, and the ensuing projection assists the gate
fidelity.  This approach is reminiscent of that used to create Fock states
by means of a sequence of conditional measurements \cite{Garraway1994}.  In
this sense the role of the measurement is quite different to the continuous
measurement schemes used in some logic gates (see e.g.\
Refs.~\cite{Pellizzari1995,Franson2004}): here it is more a helpful herald.

The gates in this paper are based on multi-photon resonances that involve
cavity mode photons distributed amongst several modes. The absence of a
photon in a mode can break the chain of resonant interaction: this is the
key to the quantum logic processes we study. To analyse the multi-photon
resonances themselves, we adapt a technique of adiabatic elimination from
atomic physics \cite{Shore1981}. However, rather than using a chain of
atomic states coupled by coherent fields, we use a chain of coupled
cavity-atom states coupled with small numbers of photons.

In the following we first briefly set up the general multi-level and
multi-mode system in Sec.\ \ref{sec:model}, and then give a simple
application to an i\textsc{swap} gate in Sec.\ \ref{sec:iswap}.  In Sec.\
\ref{sec:Fredkin}, which contains the main results of the paper, we use the
theory of effective Hamiltonians \cite{Shore1981} to describe the
multi-photon operation of a Fredkin gate.  Two Fredkin gate schemes are
presented: the first (Sec.\ \ref{subsec:FredSlow}) illustrates the basic
ideas and the second approach (presented in Sec.\ \ref{subsec:FredFast})
improves the operation of the gate.  Details of the adiabatic elimination
procedures are in the Appendices. In Sec.\ \ref{sec:rots} we find that it is
possible to realize the $x$-rotation and $z$-rotation gates, which together
with the Fredkin gate, form a universal set \cite{Nielsen2000}. The paper
concludes with Sec.\ \ref{sec:Conclusion} where we also discuss the
scalability of the proposed scheme.

\section{The multi-level model}
\label{sec:model}

At the heart of the atom-field interactions is the generic multi-level and
multi-mode Hamiltonian ($\hbar=1$)
\begin{eqnarray}
\label{equation: JCM}
H &=& 
\sum_{i,\alpha,\beta}
H_i^{\alpha\beta}
  \nonumber\\
 &=& 
\sum_{i,\alpha,\beta}
 E_\beta \Sig{\beta\beta} + E_\alpha \Sig{\alpha\alpha} 
    + \omega_i \Cre_i\Ann_i
    +\Gee{\alpha\beta}{i} \left( \Ann_i \Sig{\alpha\beta} +  \Cre_i \Sig{\beta\alpha}    \right)
  \nonumber\\
 &&
\end{eqnarray}
where a mode of frequency $\omega_i$ is coupled to two atomic levels
\ket{\alpha}, \ket{\beta}, with energies $E_\alpha$ and $E_\beta$ (with
$E_\alpha > E_\beta$).  The coupling strength is $\Gee{\alpha\beta}{i}$, and
the atomic operators $\Sig{\alpha\beta} \equiv\ket{\alpha}\bra{\beta}$.  For
multi-mode fields and multilevel atoms we will have many possibilities for
selecting the modes and atomic states.  We will assume that any given pair
of atomic levels either couples to a single cavity mode or is extremely
non-resonant, and that the Hamiltonian for the system can be written as the
sum of terms Eq.\ \eqref{equation: JCM}.

\begin{figure}[ht]
\includegraphics{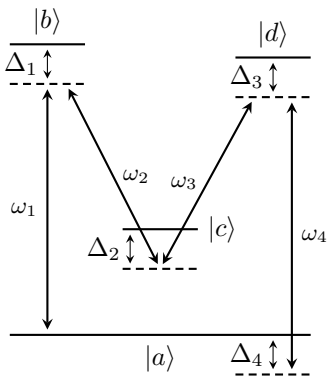}
\caption{A double-lambda scheme with four modes and four atomic levels,
  which can form an i\textsc{swap} gate for photonic qubits
  (section~\ref{sec:iswap}).  At certain interaction times this leads to a
  rearrangement of excitations between the cavity modes without a resulting
  excitation of the atom.  The detunings $\Delta_i$ represent the detuning
  of virtual states in the multi-photon resonance found when $\Delta_4 \sim
  0$.
  \label{figure2} }
\end{figure}

Before analyzing the i\textsc{swap} and Fredkin gates in detail in Secs.\
\ref{sec:iswap} and \ref{sec:Fredkin}, we start with a brief illustration of
the concepts of the swapping process with cavity excitations as depicted for
i\textsc{swap} with a double-lambda scheme as shown in Fig.\ \ref{figure2}.  Consider
the case of the initial state of the system such that the atom is in the
state \ket{a}, modes one and three both have a single excitation, and modes
two and four have no excitations (so that the overall state can be
represented by \ket{1010,a}).  Then the system can resonantly oscillate
between the initial state and the state where the excitations have been
moved to modes two and four: i.e.\ it oscillates between \ket{1010,a} and
\ket{0101,a} if $\Delta_4\rightarrow 0$. Three other states of the system,
\ket{0010,b}, \ket{0110,c} and \ket{0100,d} are also accessible. To stop
states of the atom other than \ket{a} being populated the transitions may be
detuned, i.e.  detunings $\Delta_{1,2,3}$ are large. If we have the limit
$\Delta_4\rightarrow 0$, then only the two states of interest, \ket{1010,a}
and \ket{0101,a} may be populated.  The price to be paid for detuning the
intermediate states is a slower gate, as we will see in the Section
\ref{sec:Fredkin} when we analyze the six-mode Fredkin gate.  If the two
excitations in the system are initially in either of the alternate
configurations \ket{1001,a} or \ket{0110,a} then no movement of excitations
can occur as the large detuning of the intermediate states makes this
energetically unfavorable. If we apply the encoding of a qubit, as in Table
\ref{Table: qubit}, this system can map qubits onto an i\textsc{swap} gate as seen in
detail in the next section.

\section{i\textsc{swap} gate}
\label{sec:iswap}
\subsection{Configuration for a basic i\textsc{swap} gate}

Given an atomic system of energy levels coupled to two dual rail qubits as
shown in Fig.\ \ref{figure2}, we have seen that it may be possible to limit
the system to two essential states which will oscillate. The resulting
interaction is equivalent to an i\textsc{swap} gate \cite{Schuch2003} which is shown
in Table \ref{Table:iSWAP}.

First we map four possible initial states of the system to qubits as in Table
\ref{Table: qubit}:
\begin{equation}
\begin{split}
\ket{0110,a}&\mapsto\ket{00,a}\\
\ket{0101,a}&\mapsto\ket{01,a}\\
\ket{1010,a}&\mapsto\ket{10,a}\\
\ket{1001,a}&\mapsto\ket{11,a}\,,
\end{split}
\end{equation}
where the $a$ is a reminder that the atom is always entered in state \ket{a}.
\begin{table}[t]
\begin{tabular}{ccc}
Input && Output\\
\hline
\ket{00} && \ket{00}\\
\ket{01} && i\ket{10}\\
\ket{10} && i\ket{01}\\
\ket{11} && \ket{11}
\end{tabular}
\caption{Truth table for  the i\textsc{swap} gate \cite{Schuch2003}. The i\textsc{swap} gate is locally
equivalent to a combined \textsc{cnot} and \textsc{swap} operation \cite{Schuch2003}, and forms
a universal set with the one qubit rotation gates.\label{Table:iSWAP}}
\end{table}

The Hamiltonian for this system in an interaction picture for the logical
states \ket{01} and \ket{10} is
\begin{equation}
\begin{split}
H_I = &-\Delta_1\OP{n}_1 + \Gee{ab}{1}\left(\Ann_1\Sig{ba} + \Cre_1\Sig{ab}\right)\\
&+(\Delta_2-\Delta_1)\OP{n}_2 + \Gee{bc}{2}\left(\Ann_2\Sig{bc} + \Cre_2\Sig{cb}\right)\\
&+(\Delta_2-\Delta_3)\OP{n}_3 + \Gee{cd}{3}\left(\Ann_3\Sig{dc} + \Cre_3\Sig{cd}\right)\\
&+(\Delta_4-\Delta_3)\OP{n}_4 + \Gee{da}{4}\left(\Ann_4\Sig{da} +
  \Cre_4\Sig{ad}\right) . \\
\end{split}
\label{eq:HIswapraw}
\end{equation}
The derivation of this Hamiltonian may be found in Appendix \ref{app:iswap}.
By choosing $\Delta_{1,2,3}\gg
(\Gee{ab}{1},\Gee{bc}{2},\Gee{cd}{3},\Gee{da}{4},\Delta_4)$ a two-level
effective Hamiltonian may also be derived as shown in Appendix
\ref{app:iswap}. This amounts to an adiabatic elimination of off-resonant
states. The effective Hamiltonian operates on two qubit states of the
system, i.e.\ \ket{10}\ket{a} and \ket{01}\ket{a}.  The effective
Hamiltonian takes the form
\begin{equation}
\eff{H} = \SP\SM\eff{\Delta}+\eff{g}\left(\SP+\SM\right)\,,
\end{equation}
where $\SP=\ket{01,a}\bra{10,a}$, $\SM=\ket{10,a}\bra{01,a}$ and
\begin{align}
\eff{\Delta} &\simeq \Delta_4-\frac{(\Gee{da}{4})^2}{\Delta_3}+\frac{(\Gee{ab}{1})^2}{\Delta_1}\\
\eff{g} &\simeq -\frac{\Gee{ab}{1}\Gee{bc}{2}\Gee{cd}{3}\Gee{da}{4}}{\Delta_1\Delta_2\Delta_3}\,.
\end{align}
The time evolution of the system for the states of interest is then given by
\begin{equation}
\begin{split}
\ket{10,a}&\mapsto\cos(\eff{g}t)\ket{10,a}-i\sin(\eff{g}t)\ket{01,a}\\
\ket{01,a}&\mapsto\cos(\eff{g}t)\ket{01,a}-i\sin(\eff{g}t)\ket{10,a}\,,
\end{split}
\end{equation}
and the two other logical states of the system, \ket{00} and \ket{11} are
unchanged as there are no resonant interactions and the detunings
$\Delta_{1,2,3}$ are large.  By choosing an appropriate interaction time
($|\eff{g}t|=\pi/2$) an i\textsc{swap} gate operation is realized.

Unfortunately this gate is relatively slow, as it depends on a four photon
process. Assuming that a typical detuning $\Delta_i$ should be an order of
magnitude larger than a typical coupling constant $g_j$ to make the
effective Hamiltonian a good approximation, the effective coupling constant
\eff{g} will be three orders of magnitude smaller than a typical coupling
constant. In a micromaser-like system with a coupling $g_j
/(2\pi)\approx10^4$\ Hz and a quality factor $Q\approx10^{10}$ at
$\omega/(2\pi)\approx10^{10}$\ Hz \cite{Walther2006} the gate time will only
be an order of magnitude smaller than the decay time of the cavity.

\section{Multi-photon Fredkin gate}
\label{sec:Fredkin}

\subsection{Configuration for a basic multi-photon Fredkin gate}
\label{subsec:FredSlow}

\begin{figure}[ht]
\includegraphics{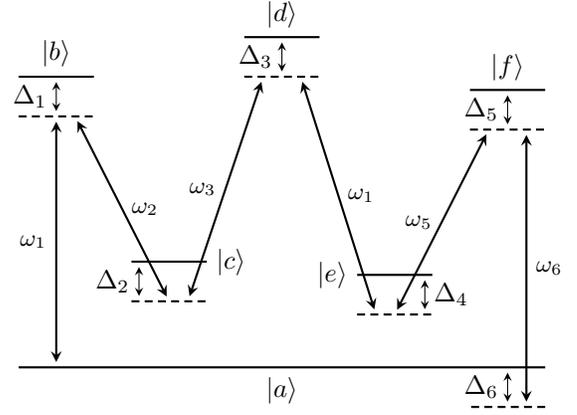}
\caption{The level scheme, with repeated mode $\omega_1$, which can realize
  a Fredkin (controlled swap) gate when $\Delta_6\sim 0$ (multiphoton
  resonance). The mode $\omega_4$ is not shown here.  By tuning $\Delta_3$
  according to the resonance conditions \eqref{3-state-res} the operational
  speed of the gate can be increased.
\label{figure3} 
}
\end{figure}

In order to build up a complete set of gates for dual-rail CQED QIP we need
a faster gate than the i\textsc{swap} gate which also entangles qubits, such as the
multi-qubit entangling Fredkin gate. To form this gate we actually add two
more transitions to the i\textsc{swap} gate configuration. This trades a four photon
process for a six photon process that will be slower, but in section
\ref{subsec:FredFast} we show that a faster gate can be produced by allowing
another state of the system to be resonant.  These extra transitions both
couple to the same mode, so that the presence of a photon in this additional
mode is required to enable the swap interaction in the remaining modes, and
its absence will prohibit the interaction. In this way we will realize a
Fredkin gate \cite{Nielsen2000,Everitt2007}.  Figure \ref{figure3} shows how
the transitions and modes can be arranged to facilitate this.  Making a
transition from \ket{a} completely around the loop of atomic states will
absorb the photon from mode one, and then return it.

To understand the full dynamics in detail, we write the Hamiltonian of the
system in an interaction picture as
\begin{equation}
\label{Equation HFred}
\begin{split}
 H_I=& -\Delta_1\OP{n}_1 + \Gee{ab}{1}\left(\Ann_1\Sig{ba}+\Cre_1\Sig{ab}\right)\\
& +(\Delta_2-\Delta_1)\OP{n}_2 + \Gee{bc}{2}\left(\Ann_2\Sig{bc}+\Cre_2\Sig{cb}\right)\\
& +(\Delta_2-\Delta_3)\OP{n}_3 + \Gee{cd}{3}\left(\Ann_3\Sig{dc}+\Cre_3\Sig{cd}\right)\\
& +\left(\OP{n}_1+\OP{n}_2+\OP{n}_3+\OP{n}_6-\Sig{aa}-\Sig{cc}\right)(\Delta_1-\Delta_3+\Delta_4)\\
& + \Gee{de}{1}\left(\Ann_1\Sig{de}+\Cre_1\Sig{ed}\right)\\
& +(\Delta_4-\Delta_5)\OP{n}_5 + \Gee{ef}{5}\left(\Ann_5\Sig{fe}+\Cre_5\Sig{ef}\right)\\
& +(\Delta_6-\Delta_5)\OP{n}_6 + \Gee{fa}{6}\left(\Ann_6\Sig{fa}+\Cre_6\Sig{af}\right)
\end{split}
\end{equation}
where $\OP{n}_i=\Cre_i \Ann_i$. Details can be found in Appendix
\ref{appa:1} where we have carefully chosen the basis to make the
Hamiltonian time independent.  An overall energy shift has been neglected,
and the detunings $\Delta_j$ are as shown in Fig.\ \ref{figure3}.

Our aim of rearranging excitations in the target modes ($\omega_2, \omega_3,
\omega_5, \omega_6$) could be achieved by having the resonant case (all
$\Delta_i = 0$), but this is a rather special case which is very sensitive
to decoherence. The optimization of the gate operation has turned out to be
an interesting problem with a number of different solutions.  Another
approach would be to aim for a multiphoton resonance: i.e.\ $\Delta_6=0$
with large detunings $\Delta_{1\mbox{--}5}$ (see Fig.\ \ref{figure3}).  We
could then use an effective two-level Hamiltonian \cite{Cook1979,Shore1981}
to model the dynamics of the whole system as if it were an atomic ladder
(see Appendix \ref{appa:1.5}).  However, an important difference here is
that the state space of the system involves both atomic states \emph{and}
quantized cavity field states.

To encode the qubits, the first mode ($\omega_1$) is paired up with a mode
($\omega_4$) that is not resonant with any transition and does not appear in
Fig.\ \ref{figure3}.  If we then label the other modes as shown in Fig.\
\ref{figure3}, we have three logical qubits represented as
\begin{equation}
\label{order}
  \ket{q_1,q_2,q_3}  \equiv   \ket{n_1n_4,n_2n_3,n_5n_6}
.
\end{equation}
Only two states of the \textit{qubit system} are effectively resonant, i.e.\
a coupling exists between the logical states
\begin{equation}
\ket{101,a}\leftrightarrow\ket{110,a}
\,,
\label{eq:Fredkin:states}
\end{equation}
and the six remaining configurations of qubits (listed in Table
\ref{table_fredkin_1}) do not make transitions. The two states in Eq.\
\eqref{eq:Fredkin:states} are coupled via five other intermediate atom-field
states.  Close to the multi-photon resonance we can produce an effective
two-level Hamiltonian \cite{Shore1981}. The derivation may be found in
Appendix \ref{appa:1.5} and the effective Hamiltonian is
\begin{equation}
\eff{H} = \SP\SM\eff{\Delta}+\eff{g}\left(\SP+\SM\right)
\label{HeffTLS}
\end{equation}
where $\hat\sigma^+ \equiv \ket{110,a}\bra{101,a}$ and the atomic state
could be factored out from the effective Hamiltonian. The effective coupling
is
\begin{equation}
\eff{g}\simeq
-
\frac{\Gee{ab}{1}\Gee{bc}{2}\Gee{cd}{3}\Gee{de}{1}\Gee{ef}{5}\Gee{fa}{6}}{\Delta_1\Delta_2\Delta_3\Delta_4\Delta_5}
\end{equation}
and the resonance condition, which has gained a component due to level
shifts, is
\begin{equation}
\label{equation: resonance}
\eff{\Delta} \simeq \Delta_6 -
\frac{\left(\Gee{fa}{6}\right)^2}{\Delta_5}
\,.
\end{equation}
We can now see that the condition for resonance is only very approximately $
\Delta_6 = 0$ and we should actually use $\eff{\Delta}=0$ which implies
$\Delta_6\simeq (\Gee{fa}{6})^2 / \Delta_5$. Thus the evolution of the
system, initially in the state \ket{101,a} or \ket{110,a}, is effectively
\begin{equation}
\begin{split}
\ket{101,a}&\mapsto\cos(\eff{g}t)\ket{101,a}-i\sin(\eff{g}t) \expup{i\eta t} \ket{110,a}\\
\ket{110,a}&\mapsto\cos(\eff{g}t)\ket{110,a}-i\sin(\eff{g}t) \expup{i\eta t}\ket{101,a}
,
\end{split}
\label{TLSevolution}
\end{equation}
where to obtain the exact Fredkin gate truth table (Table
\ref{table_fredkin_1}) without phase factors, we have undone the
$\hat\Theta_2$ transformation of Appendix~\ref{appa:1}. Then a phase factor
depending on
\begin{equation}\label{eq:eta}
\eta=\Delta_1-\Delta_3+\Delta_4\,.
\end{equation}
appears in Eq.~(\ref{TLSevolution}) which amounts to trivially changing the phase of the couplings. 
%
\begin{table}
\centering
\begin{tabular}{ccc}
Input && Output\\
\hline
\ket{000} && \ket{000}\\
\ket{001} && \ket{001}\\
\ket{010} && \ket{010}\\
\ket{011} && \ket{011}\\
\ket{100} && \ket{100}\\
\ket{110} && \ket{101}\\
\ket{101} && \ket{110}\\
\ket{111} && \ket{111}
\end{tabular}
\caption{Truth table for the effective two-level system with
  $|g_\mathrm{eff}t|=\pi/2$ and $\eta t =\pi/2$ [see Eq.\ \eqref{TLSevolution}] which
  produces a Fredkin gate \cite{Nielsen2000}.\label{table_fredkin_1}}
\end{table}
For effective operation, a typical coupling constant in Eq.\ \eqref{Equation
  HFred} must be at least an order of magnitude smaller than its associated
detuning.  This implies that the effective coupling constant \eff{g} would
be five orders of magnitude smaller than a typical cavity coupling constant,
which would leave the interaction prohibitively slow, even in modern
cavities.  We will improve this coupling constant with the approach given in
the next section.

\subsection{Configuration for a fast multi-photon Fredkin gate}
\label{subsec:FredFast}

To improve \emph{significantly} the performance of the gate we allow the
transitions to level \ket{d} to be resonant (see Fig.\ \ref{figure3}).  This
gives a useful compromise between sensitivity and an improved gate operating
time.  We extend the effective Hamiltonian method of Ref.\ \cite{Shore1981}
to a three-level case as indicated in Appendix \ref{appa:2}. Then the
conditions for resonance are:
\begin{equation}
\begin{split}
&\Delta_3 \simeq
\frac{\left(\Gee{cd}{3}\right)^2}{\Delta_2}+\frac{\left(\Gee{de}{1}\right)^2}{\Delta_4}-\frac{\left(\Gee{ab}{1}\right)^2}{\Delta_1}
\,,\\
&\Delta_6 \simeq \frac{\left(\Gee{fa}{6}\right)^2}{\Delta_5} 
\,,
\label{3-state-res}
\end{split}
\end{equation}
[see Eq.\ (\ref{3levelmatrix})].
There are now two effective coupling constants,
\begin{align}
g_1 &\simeq \frac{\Gee{ab}{1}\Gee{bc}{2}\Gee{cd}{3}}{\Delta_1\Delta_2},& g_2 &\simeq
\frac{\Gee{de}{1}\Gee{ef}{5}\Gee{fa}{6}}{\Delta_4\Delta_5}
\,,
\label{3-state-gs}
\end{align}
which can be read off from Eq.\ (\ref{3levelmatrix}).
In the case of resonance the state
\ket{110,a} evolves as
\begin{equation}
\begin{split}
\ket{110,a}  \mapsto &\phantom=\left[\bar{g}_2^2+\bar{g}_1^2\cos\left(g't\right)\right]\ket{110,a}\\
& +i\bar{g}_1\sin\left(g't\right)\ket{\phi}\\
& + \bar{g}_1\bar{g}_2\left[\cos\left(g't\right)-1\right]\expup{i\eta t}\ket{101,a},
\end{split}
\label{3-state-res_SOLN}
\end{equation}
where $\ket{\phi}$ is an auxiliary state (\ket{001010,d})
which does not have an
interpretation in our encoding of qubits. The couplings
$\bar{g}_1, \bar{g}_2$ and $g'$ are given by 
\begin{align}
g' &= \sqrt{g_1^2+g_2^2},& \bar{g}_{1,2} &= g_{1,2}/ g'
\,,
\end{align}
and for the phase factor $\eta t$, $\eta$ is as given in \eqref{eq:eta}.

To realise a Fredkin gate we need complete population transfer, so
$g't=\pi$, and if $g_1 t =g_2 t= \pi/\sqrt{2}$ and $\eta t=\pi$ then
$\ket{110,a}\mapsto\ket{101,a}$ and $\ket{101,a}\mapsto\ket{110,a}$. We then
obtain Table \ref{table_fredkin_1} without complex coefficients.  Although
we have made the assumption that the effective couplings for both
multiphoton transitions are equal ($g_1=g_2$), there is ample freedom to
tune this with the various $\Delta_i$.

\begin{figure}[ht]
\includegraphics[height=2.0in]{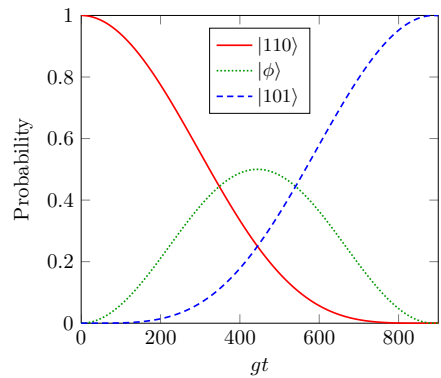}
\caption{[Color online] The Fredkin gate using the three-state model of
  Eqs.~(\ref{3-state-res}) for parameters
  in the Hamiltonian \eqref{Equation HFred}.
  The coupling
  constants $g_\mathrm{ab}$ to $g_\mathrm{fa}$ are set to
  $g$ and $\Delta_1$, $\Delta_2$, $\Delta_4$ and $\Delta_5$
  are all set to 20$g$ so that $g_1=g_2$. The detunings
  $\Delta_3$ and $\Delta_6$ are set by the resonance
  conditions \eqref{3-state-res}.
  This plot shows populations of the three states of the system which
  oscillate to produce the gate evolution [approximately given by Eq.\ 
  \eqref{3-state-res_SOLN}].
  Other states that represent
  qubits will remain essentially unchanged.
  \label{figure4}}
\end{figure}

\begin{figure}[ht]
\includegraphics[height=2.1in]{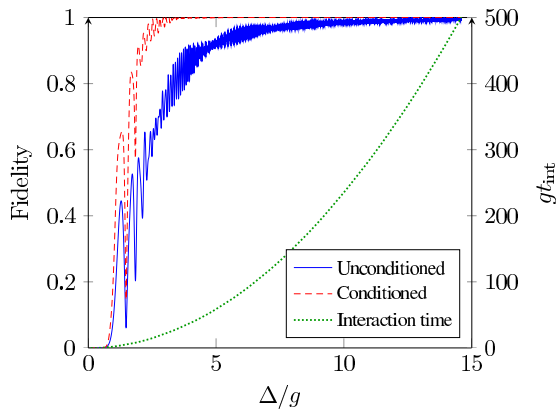}
\caption{[Color online]
Left axis:
Fidelity of the Fredkin gate (solid line) as a function of
a detuning $\Delta=\Delta_i$, for $i=1,2,4,5$.
The detunings  $\Delta_3$ and $\Delta_6$ are determined by Eqs.\
\eqref{3-state-res} with couplings $g_i^{\alpha\beta}=g$.
The dashed line shows the improvement resulting from a
conditional measurement on the atom after it has left the
cavity. 
Right axis: The dotted line shows the interaction time
$t_\mathrm{int} =\pi/g'$ as found from the three-state model
\eqref{3-state-res_SOLN}.
\label{figure5}}
\end{figure}

We have tested this theory by numerically integrating the full Hamiltonian
\eqref{Equation HFred} and checking that the appropriate gate operation
takes place.  Figure \ref{figure4} shows the population of three of the
states corresponding to those in Eq.\ \eqref{3-state-res_SOLN} with good
population swapping.  The fidelity of the exact numerical dynamics to the
analytic behavior in Eq.\ \eqref{3-state-res_SOLN} is shown in Fig.\
\ref{figure5} (solid line).  We note that the fidelity can be enhanced
(dashed line) by measuring the atom state to be \ket{a}. This forms a simple
error correction or state locking. If the atom is not found to be in state
\ket{a}, the logic operation must be aborted.

\section{Qubit rotations}
\label{sec:rots}

In addition to the Fredkin gate, an entangling multi-qubit gate, we also
need to be able to rotate an individual qubit over the Bloch-sphere to
complete a universal set of gates.  Rotations in any two of $x$, $y$ and $z$
are capable, in combination, of producing an arbitrary rotation, so it is
sufficient to show that two of these rotations are possible using the system
detailed above.  The Jaynes-Cummings model with a large detuning realizes a
simple rotation about $z$ ($\hat R_z$) with a two-level atom detuned from
the first mode of a qubit, and far detuned from the other mode.  With the
qubit represented as in Table~\ref{Table: qubit}, we will have $\ket{1}
\rightarrow \expup{i(g^2/\Delta) t}\ket{1}$ while $\ket{0} \rightarrow
\ket{0}$ at time $t$.  If the cavity field varies spatially, one can simply
adjust the interaction time to compensate \cite{Englert1998}.

\begin{figure}[ht]
\includegraphics{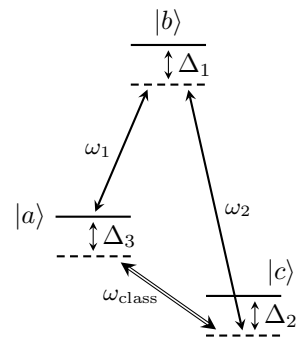}
\caption{ 
  The lambda atom has two transitions which couple to the two cavity
  modes that make up a qubit. We adiabatically eliminate levels 
  \ket{b} and \ket{c} from the interaction under the conditions
  \eqref{xrot:inequality}.
\label{figure6}}
\end{figure}

To complete the set of gates we can form an $x$-rotation using a lambda
scheme with the two transitions coupled to the two modes that make up a
qubit as shown in Fig.\ \ref{figure6}.
A strong classical field couples the ground states together with coupling
$\Omega/2$ \footnote{This may be a dipole forbidden transition, or an
  allowed two-photon transition, for example.}  and the gate is formed when
the qubit modes are detuned and a loop resonance exists.  Under the
conditions
\begin{equation}
 \Delta_1, \, \Delta_2  \gg  \Gee{ab}{1}, \, \Gee{bc}{2}, \,
 \Omega/2, \, \Delta_3
\label{xrot:inequality}
\end{equation}  
we again have an effective two-level system with the atom in state \ket{a}
in both ``levels.''  The resonance condition is (see Appendix \ref{app:xRot})
\begin{equation}\label{xRot-Detuning}
\eff{\Delta} =  \Delta_3 + \frac{(\Gee{ab}{1})^2}{\Delta_1}
\end{equation}
which
can be controlled via several free parameters. Similarly, the
effective coupling constant is
\begin{equation}\label{xRot-g}
\eff{g} = -\frac{\Gee{ab}{1} \Gee{bc}{2}\Omega}{2 \Delta_1\Delta_2},
\end{equation}
so that, finally, the  $x$-rotation operation becomes
$
\hat R_x(t) = \cos(g_\mathrm{eff}t) \hat I
            -i\sin(g_\mathrm{eff}t) \hat\sigma_x
$.

\section{Conclusion}
\label{sec:Conclusion}

In summary, we have seen that multi-photon resonances in cavities can
realize a universal set of logic gates. The multiphoton resonances are
possible because of the strong single-photon couplings (and long decay
times) available in high-$Q$ cavities.  We can locate the resonances by
utilizing effective Hamiltonians for the combined atom-field states.  For
realizable microwave cavities, we find a Fredkin gate operation time of
$4\times 10^{-4}$~s for $\Delta\sim 5g$, which is well within a photon
lifetime of $\sim0.3$~s.
Results on the more quantitative effects of decoherence are discussed in
Ref.~\cite{Everitt14}.
If we make a measurement-based selection of the atom leaving the cavity, the
fidelity is $\sim1-6\times 10^{-4}$ (at $\Delta \sim 5 g$ in Fig.\
\ref{figure5}), falling to $\sim0.91$ if no measurement is made.  Some
tuning of the energy levels may be achieved by Stark shifting the Rydberg
states \cite{Larson2004}.

The system is reasonably insensitive to variations in parameters (Fig.\
\ref{figure5}).  A Fredkin gate based on resonance, i.e.\ with all the
$\Delta_i=0$ (Fig.\ \ref{figure3}), would be very sensitive and require
specific coupling constants \cite{Cook1979}. By allowing resonance in just a
few places, i.e.\ with $\Delta_3$ and $\Delta_6$, we have reached a
practical compromise on sensitivity and gate speed.

\begin{figure}[ht]
\includegraphics[width=0.45\textwidth]{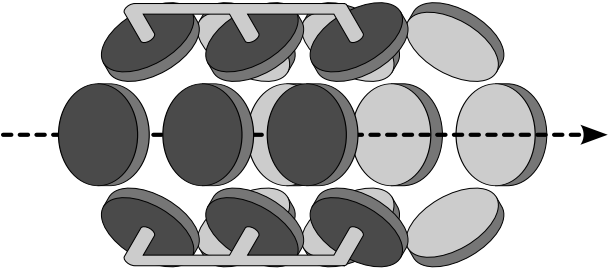}
\caption{
  Conceptual, and simplified, illustration of cascaded clusters of six cavities resulting
  in a scalable system. The path of the ancilla atom is indicated
  with the arrow. Not shown are the electrodes to be used locally to Stark shift
  atomic levels out of resonance with the cavity modes as required for the
  implemented gates.
  \label{figure7}}
\end{figure}

The system appears to be scalable, though in this paper we have focused on
basic logic gates.  Of course, scalability is a crucial issue for building a
QIP architecture and we indicate some ways in which this can be done. Figure
\ref{figure7} shows the path of an atom through a cascade of cavity
clusters, each involving six modes. In this case, depending on whether the
levels are in the configuration of Fig. \ref{figure3}, or Fig.
\ref{figure6}, we can have different gates operating between the photonic
qubits. We recall that in each cluster the atom acts \emph{only as an
  ancilla} to bring about the operation. To facilitate communication between
cavity clusters, one pair of modes could form a \emph{bus mode}, if oriented
along the axis of travel of the atom. Such a mode pair could interact with
any of the cavity clusters as the ancilla atom passes through.
Alternatively, the qubit state could be temporarily transferred to the atom
to allow inter-cluster communication (though in such a case atoms would have
to fly through the cavities in both directions to ensure a two-way flow of
information). In all cases, a simple local Stark detuning could be used to
``turn off'' a cluster, i.e.\ to prevent its interaction with the ancilla,
or bus-mode. Through all these methods, we believe that more complex gates
could be built up. However, the multi-photon cavity resonances at the heart
of the gates appear to be in the range of practicality. They have not been
observed to date and may be an interesting phenomena in themselves with
other applications such as the creation of entangled states in cavity
resonators.

We thank the UK EPSRC, the Leverhulme Trust and the Japanese Society for the
Promotion of Science for their support. Special thanks to Moteb M Alqahtani
for comments on the manuscript and to B.~Shore.

\appendix
\section{Derivation of the i\textsc{swap} Gate Effective Hamiltonian}\label{app:iswap}
\subsection{Transformation}
The Hamiltonian that describes the level scheme in Fig.\ \ref{figure2} in the
Schr\"odinger picture is
\begin{multline}
H = \sum_\alpha E_\alpha\Sig{\alpha\alpha} + \sum_{i=1}^4\omega_i\Cre_i\Ann_i
+\left[\Gee{ab}{1}\Sig{ba}\Ann_1+\Gee{bc}{2}\Sig{cb}\Cre_2\right.\\
\left.+\Gee{cd}{3}\Sig{dc}\Ann_3+\Gee{da}{4}\Sig{ad}\Cre_4 + \mathrm{h.c.}\right]\,,
\end{multline}
where $\alpha$ represents the energy levels of the atom and $i$ the modes of
the field. A transformation operator $T=\exp(i\OP{\Theta} t)$ is defined to
move to an interaction picture and remove absolute energy dependence. The
new Hamiltonian is given by
\begin{equation}\label{transformation_operator}
H^\prime=\OP{T}H\OP{T}^\dagger-\OP{\Theta}
\end{equation}
The choice of the operator $\Theta$ is made to remove the non-interacting
terms in the Hamiltonian without introducing time dependence. That is, we
let
\begin{equation}
\Theta=\sum_\alpha E_\alpha\Sig{\alpha\alpha}+\sum_{i=1}^4\Cre_i\Ann_i(\omega_i+\delta_i)\,,
\end{equation}
where the $\delta_i$ are related to the $\Delta_j$ in the level diagram Fig.\
\ref{figure2} by
\begin{equation}
\begin{split}
\Delta_1 &=\delta_1\\
\Delta_2 &=\delta_1-\delta_2\\
\Delta_3 &=\delta_1-\delta_2+\delta_3\\
\Delta_4 &=\delta_1-\delta_2+\delta_3-\delta_4\,,
\label{eq:bigdsmalld}
\end{split}
\end{equation}
and $\delta_i$ are interpreted as the detuning between a pair of levels and
a field mode, whereas $\Delta_j$ are the detunings of levels from level
\ket{a} and a multi-photon transition.  This transformation yields the
Hamiltonian
\begin{multline}
H = -\sum_{i=1}^4\delta_i\Cre_i\Ann_i
+\left[\Gee{ab}{1}\Sig{ba}\Ann_1+\Gee{bc}{2}\Sig{cb}\Cre_2\right.\\
\left.+\Gee{cd}{3}\Sig{dc}\Ann_3+\Gee{da}{4}\Sig{ad}\Cre_4 + \mathrm{h.c.}\right]\,,
\end{multline}
which leads us to Eq.\ \eqref{eq:HIswapraw} on utilizing Eq.\
\eqref{eq:bigdsmalld}
for the $\delta_i$.

\subsection{The Effective Hamiltonian}\label{app:effective_how_to}
We follow a procedure from a paper by Shore \cite{Shore1981}. We define two
projection operators, $\OP{P}$ and $\OP{Q}$, to select out the states close
to resonance and far from resonance, respectively. In this assumption
$\Delta_i\gg\Gee{\alpha\beta}{j}$ where \Gee{\alpha\beta}{j} represents all
of the coupling constants and $i=1,2,3,4$. $\Delta_4$ will be small and
chosen later to ensure resonance. From this we define the components $H_0
=\OP{P}H\OP{P}^\dagger$, $\OP{B}=\OP{P}H\OP{Q}^\dagger$ and
$A=\OP{Q}H\OP{Q}^\dagger$. For systems with a small Hilbert space it is
simplest to proceed in matrix form. If the system is initially in one of the
states \ket{1010,a} or \ket{0101,a} then at some time later the state of the
system will be
$\ket{\Psi}=c_1\ket{1010,a}+c_2\ket{0010,b}+c_3\ket{0110,c}+c_4\ket{0100,d}+c_5\ket{0101,a}$.
The Hamiltonian in matrix form is
\begin{align}
&H=\begin{pmatrix}
0 & \Gee{ab}{1} & 0 & 0 & 0\\
\Gee{ab}{1} & \Delta_1 & \Gee{bc}{2} & 0 & 0\\
0 & \Gee{bc}{2} & \Delta_2 & \Gee{cd}{3} & 0\\
0 & 0 & \Gee{cd}{3} & \Delta_3 & \Gee{da}{4}\\
0 & 0 & 0 & \Gee{da}{4} & \Delta_4
\end{pmatrix}\,,
&\ket{\Psi}=\begin{pmatrix}c_1\\c_2\\c_3\\c_4\\c_5\end{pmatrix}\,.
\end{align}
The operators $H_0$, $\OP{A}$ and $\OP{B}$ can be found by manipulating the Hamiltonian
matrix so that the two states close to resonance are put to the top of the state vector.
\begin{align}
&H=\begin{pmatrix}
0 & 0 & \Gee{ab}{1} & 0 & 0\\
0 & \Delta_4 & 0 & 0 & \Gee{da}{4}\\
\Gee{ab}{1} & 0 & \Delta_1 & \Gee{bc}{2} & 0\\
0 & 0 & \Gee{bc}{2} & \Delta_2 & \Gee{cd}{3}\\
0 & \Gee{da}{4} & 0 & \Gee{cd}{3} & \Delta_3
\end{pmatrix}\,,
&\ket{\Psi}=\begin{pmatrix}c_1\\c_5\\c_2\\c_3\\c_4\end{pmatrix}\,.
\end{align}
When this has been done the Hamiltonian is broken into the parts.
\begin{align}
&H_0=\begin{pmatrix}0&0\\0&\Delta_4\end{pmatrix}\,,
&A=\begin{pmatrix}
\Delta_1 & \Gee{bc}{2} & 0\\
\Gee{bc}{2} & \Delta_2 & \Gee{cd}{3}\\
0 & \Gee{cd}{3} & \Delta_3
\end{pmatrix}\,,
\end{align}
\begin{equation*}
B = \begin{pmatrix}
\Gee{ab}{1} & 0 & 0\\
0 & 0 & \Gee{da}{4}
\end{pmatrix}\,.
\end{equation*}
Using these parts the effective Hamiltonian is constructed according to the
equation \cite{Shore1981}
\begin{equation}
\label{eq:Heff:basic}
\eff{H}=H_0-BA^{-1}B^\dagger\,,
\end{equation}
which, after a trivial energy shift, results in the two-state effective
Hamiltonian
\begin{equation}
\eff{H}=\begin{pmatrix}
0 & \eff{g}\\
\eff{g} & \eff{\Delta}
\end{pmatrix}\,,
\end{equation}
where
\begin{equation}
\begin{split}
\eff{g}&=-\frac{\Gee{ab}{1}\Gee{bc}{2}\Gee{cd}{3}\Gee{da}{4}}{\Delta_1\Delta_2\Delta_3-\Delta_3(\Gee{bc}{2})^2-\Delta_1(\Gee{cd}{3})^2}\\
&\approx-\frac{\Gee{ab}{1}\Gee{bc}{2}\Gee{cd}{3}\Gee{da}{4}}{\Delta_1\Delta_2\Delta_3}\,,
\end{split}
\end{equation}
and
\begin{equation}\label{iSWAPDeff}
\begin{split}
%
%
&\eff{\Delta}=\\
&\Delta_4+\frac{
	(\Gee{ab}{1})^2(\Delta_2\Delta_3-(\Gee{cd}{3})^2)-(\Gee{da}{4})^2(\Delta_1\Delta_2-(\Gee{bc}{2})^2)}
{\Delta_1\Delta_2\Delta_3-\Delta_3(\Gee{bc}{2})^2-\Delta_1(\Gee{cd}{3})^2}\\
&\approx\Delta_4 + \frac{(\Gee{ab}{1})^2}{\Delta_1}-\frac{(\Gee{da}{4})^2}{\Delta_3}\,,
\end{split}
\end{equation}
which are the results given in Section \ref{sec:iswap}.

\section{Derivation of the Fredkin Gate Effective Hamiltonian}\label{derivation}
\subsection{Transformations}\label{appa:1}
\begin{figure}
\includegraphics{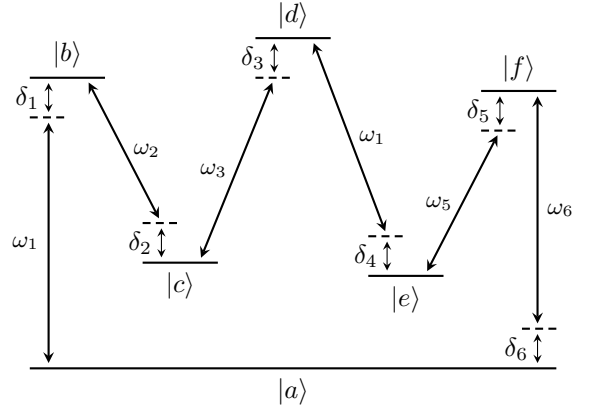}
\caption{This diagram illustrates the difference between the detunings
  $\Delta_i$ in Figure \ref{figure3} and the detunings $\delta_i$ utilized
  in Appendix \ref{derivation}. The detunings $\delta_i$ indicate the
  detuning of each coupled field from its respective transition. The
  detunings $\Delta_i$ represent the accumulated detuning of a multi-photon
  resonance (without considering level shifts).\label{figure8}}
\end{figure}

The Hamiltonian that describes the level scheme in Fig.~\ref{figure3} is
\begin{equation}
\begin{split}
H=&\sum_\alpha E_\alpha\Sig{\alpha\alpha}+\sum_{i\ne 4}\omega_i\Cre_i\Ann_i\\
&+ \left[ \Gee{ab}{1}\Sig{ba}\Ann_1 + \Gee{bc}{2}\Sig{cb}\Cre_2 +\Gee{cd}{3}\Sig{dc}\Ann_3 \right.\\
&\left.  + \Gee{de}{1}\Sig{ed}\Cre_1
+ \Gee{ef}{5}\Sig{fe}\Ann_{5} + \Gee{af}{6}\Sig{af}\Cre_6 +\mathrm{h.c.}\right]\,,
\end{split}
\end{equation}
where we note the repeated mode~1 in the interaction terms and the absence
of mode~4 so that for the sum over $i$ we have $i=1, 2, 3, 5, 6$.  To
proceed to an interaction picture, we define a transformation through
$\OP{T_1}=\exp(i\OP{\Theta}_1 t)$ which will modify the Hamiltonian
according to \eqref{transformation_operator}.
The full transformation will be made in two steps.
For the first step in the transformation we remove explicit dependence
on the atomic energy levels
\begin{equation}
\OP{\Theta}_1=\sum_\alpha E_\alpha\Sig{\alpha\alpha}+\sum_{i\ne
  4}\Cre_i\Ann_i(\omega_i+\delta_i)
\label{eq:Theta_1}
\end{equation}
where $\delta_i$ are the detunings between particular atomic transitions and
the relevant mode, i.e. $E_\mathrm{d}-E_\mathrm{c}=\omega_3+\delta_3$.
Figure \ref{figure8} shows how these relate to the $\Delta_i$ in Figure
\ref{figure3}.
Specifically
\begin{equation}
\begin{split}
\Delta_1 &=\delta_1\\
\Delta_2 &=\delta_1-\delta_2\\
\Delta_3 &=\delta_1-\delta_2+\delta_3\\
\Delta_4 &=\delta_1-\delta_2+\delta_3-\delta_4\\
\Delta_5 &=\delta_1-\delta_2+\delta_3-\delta_4+\delta_5\\
\Delta_6 &=\delta_1-\delta_2+\delta_3-\delta_4+\delta_5-\delta_6\,.
\end{split}
\end{equation}
Note that there is no provision for $\delta_4$ in the transformation with
$\hat\Theta_1$, Eq.\ \eqref{eq:Theta_1}, as mode four is not present. (It is
replaced by mode~1.) This choice for $\hat\Theta_1$ avoids time dependence
in most elements in the resultant Hamiltonian
\begin{multline}
H^\prime=-\sum_{i\ne 4}\delta_i\Cre_i\Ann_i + \left[ \Gee{ab}{1}\Sig{ba}\Ann_1 + \Gee{bc}{2}\Sig{cb}\Cre_2 +\Gee{cd}{3}\Sig{dc}\Ann_3\right.\\
\left.  + \Gee{de}{1}\expup{i(\delta_1-\delta_4)t}\Sig{ed}\Cre_1 + \Gee{ef}{5}\Sig{fe}\Ann_{5} + \Gee{af}{6}\Sig{af}\Cre_6 +\mathrm{h.c.}\right]\,.
\end{multline}
We make a second transformation to remove some remaining
time dependence
\begin{equation}\label{eq:Theta_2}
\OP{\Theta}_2 = \left(\OP{n}_1+\OP{n}_2+\OP{n}_3+\OP{n}_6-\Sig{aa}-\Sig{cc}\right)(\delta_4-\delta_1)\,.
\end{equation}
The resultant Hamiltonian is
\begin{multline}
H^{\prime\prime} = - \sum_{i\ne 4}\delta_i\Cre_i\Ann_i + \left[ \Gee{ab}{1}\Sig{ba}\Ann_1 + \Gee{bc}{2}\Sig{cb}\Cre_2 + \Gee{cd}{3}\Sig{dc}\Ann_3\right.\\
\left. + \Gee{de}{1}\Sig{ed}\Cre_1 + \Gee{ef}{5}\Sig{fe}\Ann_{5} + \Gee{af}{6}\Sig{af}\Cre_6 +\mathrm{h.c.}\right] - \OP{\Theta}_2 \,.
\end{multline}
When we express $H^{\prime\prime}$ in terms of $\Delta_i$, Eq.\ \eqref{Equation HFred} is recovered.

\subsection{Effective 2-State Behavior} \label{appa:1.5}

Following the procedure outlined in Appendix \ref{app:effective_how_to}
\cite{Shore1981} we define two projection operators, $\OP{P}$ and $\OP{Q}$,
to select out the states close to resonance and far from resonance
respectively. In this assumption $\Delta_i\gg\Gee{\alpha\beta}{j}$ where
\Gee{\alpha\beta}{j} represents all of the coupling constants and
$i=1,2,3,4,5$. $\Delta_6$ will be small and is chosen later to ensure
resonance. As before, we define the components $H_0 =\OP{P}H\OP{P}^\dagger$,
$\OP{B}=\OP{P}H\OP{Q}^\dagger$ and $A=\OP{Q}H\OP{Q}^\dagger$. Given these
new operators the effective Hamiltonian is defined as
\begin{equation}
\label{reduction}
H_{\mathrm{eff}}=H_0 - \OP{B}\OP{A}^{-1}\OP{B}^\dagger\,
\end{equation}
as in Eq.\ \eqref{eq:Heff:basic}.  The operator $\OP{P}$ projects onto
states of the system we expect to be populated, \ket{10,01,10,a} and
\ket{10,10,01,a}. The operator $\OP{Q}$ projects onto the states that we
expect to be suppressed: \ket{00,01,10,b}, \ket{00,11,10,c},
\ket{00,10,10,d}, \ket{10,10,10,e}, \ket{10,10,00,f}, \ket{00,10,01,b} and
\ket{01,10,01,c}, with this order chosen such that the matrix $A$ is
tridiagonal.
As the system has a small Hilbert space, the resultant operators are best
shown as matrices. Note that an additional trivial transformation has been
made to set the energy of the state \ket{10,01,10,a} (see Eq.~\eqref{order})
to zero as a reference point. Then
\begin{equation}
H_0=\begin{pmatrix}0&0\\0&\Delta_6\end{pmatrix}\,,\ 
B=\begin{pmatrix}\Gee{ab}{1}&0&0&0&0&0&0\\0&0&0&0&\Gee{fa}{6}&\Gee{ab}{1}&0\end{pmatrix}\,,
\end{equation}
and $A$ is a $7\times7$ matrix composed of the detuned portion of the Hamiltonian
\begin{equation}\label{FredkinA2}
A=\begin{pmatrix}
\Delta_1 & \Gee{bc}{2} & 0 & 0 & 0 & 0 & 0\\
\Gee{bc}{2} & \Delta_2 & \Gee{cd}{3} & 0 & 0 & 0 & 0\\
0 & \Gee{cd}{3} & \Delta_3 & \Gee{de}{1} & 0 & 0 & 0\\
0 & 0 & \Gee{de}{1} & \Delta_4 & \Gee{ef}{5} & 0 & 0\\
0 & 0 & 0 & \Gee{ef}{5} & \Delta_5 & 0 & 0\\
0 & 0 & 0 & 0 & 0 & \Delta_6+\Delta_1 & \Gee{bc}{2}\sqrt{2}\\
0 & 0 & 0 & 0 & 0 & \Gee{bc}{2}\sqrt{2} & \Delta_6+\Delta_1
\end{pmatrix}\,.
\end{equation}
The matrix $A$ must be inverted, but as B has two populated elements, only
four elements of $A^{-1}$ need be calculated for use in equation
\eqref{reduction}. The resultant two state effective Hamiltonian produced
using equation \eqref{reduction} is
\begin{equation}
\eff{H}=\begin{pmatrix}
0 & \eff{g}\\
\eff{g} & \Delta_{\mathrm{eff}}
\end{pmatrix}\,,
\end{equation}
where
\begin{equation}\label{FredkinDeff}
\begin{split}
g_{\mathrm{eff}} &\approx-\frac{\Gee{ab}{1}\Gee{bc}{2}\Gee{cd}{3}\Gee{de}{1}\Gee{ef}{5}\Gee{fa}{6}}{\Delta_1\Delta_2\Delta_3\Delta_4\Delta_5}\,,\\
\Delta_{\mathrm{eff}} &\approx \Delta_6 
 -\frac{(\Gee{fa}{6})^2}{\Delta_5}
\,.
\end{split}
\end{equation}

Note that \eqref{FredkinDeff} may appear to be missing a term when compared
with \eqref{iSWAPDeff}. This is due to the lower right $2\times2$ submatrix
in \eqref{FredkinA2}, which deals with a mode with two excitations. This is
not present in the i\textsc{swap} gate, and to first order leads to one less term in
\eqref{FredkinDeff}.

\subsection{Effective 3-State Behavior} \label{appa:2}

To produce an effective three level system with the states \ket{10,01,10,a},
\ket{00,10,10,d} and \ket{10,10,01,a} we generalize the procedure in
Appendix \ref{appa:1.5}. We use the same full set of states
\ket{10,01,10,a}, \ket{00,01,10,b}, \ket{00,11,10,c}, \ket{00,10,10,d},
\ket{10,10,10,e}, \ket{10,10,00,f}, \ket{10,10,01,a}, \ket{00,10,01,b} and
\ket{01,10,01,c}, allowing the state \ket{00,10,10,d} to be close to
resonance in addition to \ket{10,01,10,a} and \ket{10,10,01,a}. The detuning
$\Delta_3$ is chosen later to ensure resonance.
The resultant operators are
\begin{equation}
H_0=\begin{pmatrix}0&0&0\\0&\Delta_3&0\\0&0&\Delta_6\end{pmatrix},\  
B = \begin{pmatrix}\Gee{ab}{1}&0&0&0&0&0\\0&\Gee{cd}{3}&\Gee{de}{1}&0&0&0\\0&0&0&\Gee{af}{6}&\Gee{ab}{1}&0\end{pmatrix},
\end{equation}
and $A$ is the $6\times6$ matrix composed of the detuned portion of the Hamiltonian
\begin{equation}
A=\begin{pmatrix}
\Delta_1 & \Gee{bc}{2} & 0 & 0 & 0 & 0\\ 
\Gee{bc}{2} & \Delta_2 & 0 & 0 & 0 & 0\\
0 & 0 & \Delta_4 & \Gee{ef}{5} & 0 & 0\\
0 & 0 & \Gee{ef}{5} & \Delta_5 & 0 & 0\\
0 & 0 & 0 & 0 & \Delta_6+\Delta_1 & \Gee{bc}{2}\sqrt{2}\\
0 & 0 & 0 & 0 & \Gee{bc}{2}\sqrt{2} & \Delta_6+\Delta_2
\end{pmatrix}\,.
\end{equation}

The subsystem associated with $H_0$ is composed of the states
\ket{10,01,10,a}, \ket{00,10,10,d} and \ket{10,10,01,a}.  Then by utilizing
Eq.~\eqref{reduction}, the full matrix for the effective Hamiltonian is
\begin{equation}
\label{3levelmatrix}
H_{\mathrm{eff}}=\begin{pmatrix}
0 & g_1 & 0\\
g_1 & \Delta_1^{\mathrm{eff}} & g_2\\
0 & g_2 & \Delta_2^{\mathrm{eff}}
\end{pmatrix}\,,
\end{equation}
where
\begin{align}
g_1 &= \frac{\Gee{ab}{1}\Gee{bc}{2}\Gee{cd}{3}}{\Delta_1\Delta_2},& g_2 & =
\frac{\Gee{de}{1}\Gee{ef}{5}\Gee{fa}{6}}{\Delta_4\Delta_5}
\,,
\end{align}
and
\begin{equation}
\begin{split}
&\Delta_1^{\mathrm{eff}}\approx\Delta_3 + \frac{\left(\Gee{ab}{1}\right)^2}{\Delta_1} - \frac{\left(\Gee{cd}{3}\right)^2}{\Delta_2}-\frac{\left(\Gee{de}{1}\right)^2}{\Delta_4}\,,\\
&\Delta_2^{\mathrm{eff}}\approx\Delta_6 - \frac{\left(\Gee{af}{6}\right)^2}{\Delta_5}\,.
\end{split}
\end{equation}

The conditions for the fast Fredkin gate \eqref{3-state-res},
\eqref{3-state-gs} and \eqref{3-state-res_SOLN} are derived from the
Hamiltonian in \eqref{3levelmatrix}.  For the gate to operate, the second
and third diagonal elements must be equal to the first.

\section{Derivation of the $x$-Rotation Gate}\label{app:xRot}
The Hamiltonian that describes the system shown in Fig.~\ref{figure6} is
\begin{equation}
\begin{split}
H=&E_\mathrm{a}\Sig{aa}+E_\mathrm{b}\Sig{bb}+E_\mathrm{c}\Sig{cc}+\omega_1\Cre_1\Ann_1+\omega_2\Cre_2\Ann_2\\
&+\left[\Gee{ab}{1}\Cre_1\Sig{ab}+\Gee{bc}{2}\Cre_2\Sig{cb}+\frac{\Omega}{2}\expup{i\omega_3t}\Sig{ca}+\mathrm{h.c.}\right]\,.
\end{split}
\end{equation}
As with the derivation of the Fredkin gate, the detuning is defined for each
transition such that
\begin{equation}
\begin{split}
E_\mathrm{b}-E_\mathrm{a}&=\omega_1+\delta_1\\
E_\mathrm{b}-E_\mathrm{c}&=\omega_2+\delta_2\\
E_\mathrm{a}-E_\mathrm{c}&=\omega_3+\delta_3
\end{split}
\end{equation}
and the link between $\Delta_i$ and $\delta_i$ is
\begin{equation}
\begin{split}
\Delta_1&=\delta_1\\
\Delta_2&=\delta_1-\delta_2\\
\Delta_3&=\delta_1-\delta_2+\delta_3\,.
\end{split}
\end{equation}
As in Appendix \ref{appa:1} a transformation operator is chosen
\begin{equation}
\OP{T}=\exp\left(i\OP{\Theta}_{1}t\right)\,,
\end{equation}
where
\begin{equation}
\OP{\Theta}_{1}=\sum_{\alpha=\mathrm{a},\mathrm{b},\mathrm{c}}E_\alpha\Sig{\alpha\alpha}
+\sum_{i=1,2}\Cre_i\Ann_i\left(\omega_i+\delta_i\right)\,.
\end{equation}
After the transformation the Hamiltonian is
\begin{equation}
\begin{split}
H^\prime=&-\delta_1\Cre_1\Ann_1-\delta_2\Cre_2\Ann_2\\
&+\left[\Gee{ab}{1}\Cre_1\Sig{ab}+\Gee{bc}{2}\Cre_2\Sig{cb}+\frac{\Omega}{2}\expup{-i\delta_3t}\Sig{ca}+\mathrm{h.c.}\right] .
\end{split}
\end{equation}
A second transformation is made to remove the remaining time dependence
\begin{equation}
\OP{\Theta}_{2}=\frac{1}{2}\left(\Cre_1\Ann_1-\Cre_2\Ann_2+\Sig{cc}-\Sig{aa}\right)\delta_3-\delta_1\,.
\end{equation}
The Hamiltonian is now
\begin{equation}
\begin{split}
H^{\prime\prime}=&-\delta_1\Cre_1\Ann_1-\delta_2\Cre_2\Ann_2-\OP{\Theta}_{2}\\
&+\left[\Gee{ab}{1}\Cre_1\Sig{ab}+\Gee{bc}{2}\Cre_2\Sig{cb}+\frac{\Omega}{2}\Sig{ca}+\mathrm{h.c.}\right] .
\end{split}
\end{equation}
Assuming that only one excitation exists in the system, the wavefunction is
\begin{multline}
\ket{\Psi}=c_0\ket{1,0,c}+c_1\ket{1,0,a}\\
+c_2\ket{0,0,b}+c_3\ket{0,1,c}+c_4\ket{0,1,a}\,.
\end{multline}
The procedure for producing an effective Hamiltonian is now followed exactly
as in Ref.\ \cite{Shore1981} for atomic states alone.
The Hamiltonian, under the assumption of only one excitation, can be displayed as the matrix
\begin{align}
&H^{\prime\prime}=\begin{pmatrix}
\Delta_2-\Delta_3 & \Omega/2 & 0 & 0 & 0\\
\Omega/2 & 0 & \Gee{ab}{1} & 0 & 0\\
0 & \Gee{ab}{1} & \Delta_1 & \Gee{bc}{2} & 0\\
0 & 0 & \Gee{bc}{2} & \Delta_2 & \Omega/2\\
0 & 0 & 0 & \Omega/2 & \Delta_3
\end{pmatrix},
&\ket{\Psi}=\begin{pmatrix}c_0\\c_1\\c_2\\c_3\\c_4\end{pmatrix}.
\end{align}
The matrix is rearranged to place the two states of the system that are close to resonance to the top of the state vector. We consider the case when the atom is initially in the state \ket{a}, so both states of the system with the atom state as \ket{a} are close to resonance. In other words, $\Gee{ab}{1},\Gee{bc}{2},\Omega/2,\Delta_3\ll\Delta_1,\Delta_2$.
\begin{align}
&H^{\prime\prime}=\begin{pmatrix}
0 & 0 & \Gee{ab}{1} & 0 & \Omega/2\\
0 & \Delta_3 & 0 & \Omega/2 & 0\\
\Gee{ab}{1} & 0 & \Delta_1 & \Gee{bc}{2} & 0\\
0 & \Omega/2 & \Gee{ab}{2} & \Delta_2 & 0\\
\Omega/2 & 0 & 0 & 0 & \Delta_2-\Delta_3
\end{pmatrix},
&\ket{\Psi}=\begin{pmatrix}c_1\\c_4\\c_2\\c_3\\c_0\end{pmatrix}.
\end{align}
The top left $2\times2$ matrix is $H_0$, 
the bottom right $3\times3$ is $A$, and
the top right partition is $B$:
\begin{align}
&H_0=\begin{pmatrix}0&0\\0&\Delta_3\end{pmatrix}\,, 
&A=\begin{pmatrix}
\Delta_1&\Gee{bc}{2}&0\\
\Gee{bc}{2}&\Delta_2&0\\
0&0&\Delta_2-\Delta_3
\end{pmatrix}\,,
\end{align}
\begin{equation}
B=\begin{pmatrix}\Gee{ab}{1}&0&\Omega/2\\0&\Omega/2&0\end{pmatrix}.
\end{equation}
The assumption that
$\Gee{ab}{1},\Gee{bc}{2},\Omega/2,\Delta_3\ll\Delta_1,\Delta_2$ was already
made, so using equation \eqref{reduction} the approximate effective
Hamiltonian is
\begin{equation}
H_{\mathrm{eff}}\approx\begin{pmatrix}
-(\Gee{ab}{1})^2/\Delta_1-\Omega^2/4\Delta_2 & \Gee{ab}{1}\Gee{bc}{2}\Omega/2\Delta_1\Delta_2\\
\Gee{ab}{1}\Gee{bc}{2}\Omega/2\Delta_1\Delta_2 & \Delta_3-\Omega^2/4\Delta_2
\end{pmatrix}\,.
\end{equation}
From this equation the effective detuning is the difference between the
diagonal elements \eqref{xRot-Detuning} and the effective coupling constant
is the off-diagonal element \eqref{xRot-g}
\begin{align}
&\Delta_{\mathrm{eff}}\approx\Delta_3+\frac{(\Gee{ab}{1})^2}{\Delta_1}\,,&g_{\mathrm{eff}}\approx\frac{\Gee{ab}{1}\Gee{bc}{2}\Omega}{2\Delta_1\Delta_2}\,.
\end{align}

\bibliographystyle{apsrev4-1}
\bibliography{bibliography}

\begin{thebibliography}{48}%
\makeatletter
\providecommand \@ifxundefined [1]{%
 \@ifx{#1\undefined}
}%
\providecommand \@ifnum [1]{%
 \ifnum #1\expandafter \@firstoftwo
 \else \expandafter \@secondoftwo
 \fi
}%
\providecommand \@ifx [1]{%
 \ifx #1\expandafter \@firstoftwo
 \else \expandafter \@secondoftwo
 \fi
}%
\providecommand \natexlab [1]{#1}%
\providecommand \enquote  [1]{``#1''}%
\providecommand \bibnamefont  [1]{#1}%
\providecommand \bibfnamefont [1]{#1}%
\providecommand \citenamefont [1]{#1}%
\providecommand \href@noop [0]{\@secondoftwo}%
\providecommand \href [0]{\begingroup \@sanitize@url \@href}%
\providecommand \@href[1]{\@@startlink{#1}\@@href}%
\providecommand \@@href[1]{\endgroup#1\@@endlink}%
\providecommand \@sanitize@url [0]{\catcode `\\12\catcode `\$12\catcode
  `\&12\catcode `\#12\catcode `\^12\catcode `\_12\catcode `\%12\relax}%
\providecommand \@@startlink[1]{}%
\providecommand \@@endlink[0]{}%
\providecommand \url  [0]{\begingroup\@sanitize@url \@url }%
\providecommand \@url [1]{\endgroup\@href {#1}{\urlprefix }}%
\providecommand \urlprefix  [0]{URL }%
\providecommand \Eprint [0]{\href }%
\providecommand \doibase [0]{http://dx.doi.org/}%
\providecommand \selectlanguage [0]{\@gobble}%
\providecommand \bibinfo  [0]{\@secondoftwo}%
\providecommand \bibfield  [0]{\@secondoftwo}%
\providecommand \translation [1]{[#1]}%
\providecommand \BibitemOpen [0]{}%
\providecommand \bibitemStop [0]{}%
\providecommand \bibitemNoStop [0]{.\EOS\space}%
\providecommand \EOS [0]{\spacefactor3000\relax}%
\providecommand \BibitemShut  [1]{\csname bibitem#1\endcsname}%
\let\auto@bib@innerbib\@empty
\bibitem [{\citenamefont {Nielsen}\ and\ \citenamefont
  {Chuang}(2000)}]{Nielsen2000}%
  \BibitemOpen
  \bibfield  {author} {\bibinfo {author} {\bibfnamefont {M.~A.}\ \bibnamefont
  {Nielsen}}\ and\ \bibinfo {author} {\bibfnamefont {I.~L.}\ \bibnamefont
  {Chuang}},\ }\href
  {http://www.cambridge.org/uk/catalogue/catalogue.asp?isbn=9780521635035}
  {\emph {\bibinfo {title} {{{Quantum Computation and Quantum Information}}}}}\
  (\bibinfo  {publisher} {Cambridge University Press},\ \bibinfo {year}
  {2000})\BibitemShut {NoStop}%
\bibitem [{\citenamefont {Raimond}\ \emph {et~al.}(2001)\citenamefont
  {Raimond}, \citenamefont {Brune},\ and\ \citenamefont
  {Haroche}}]{Raimond2001}%
  \BibitemOpen
  \bibfield  {author} {\bibinfo {author} {\bibfnamefont {J.~M.}\ \bibnamefont
  {Raimond}}, \bibinfo {author} {\bibfnamefont {M.}~\bibnamefont {Brune}}, \
  and\ \bibinfo {author} {\bibfnamefont {S.}~\bibnamefont {Haroche}},\ }\href
  {\doibase 10.1103/RevModPhys.73.565} {\bibfield  {journal} {\bibinfo
  {journal} {Reviews of Modern Physics}\ }\textbf {\bibinfo {volume} {73}},\
  \bibinfo {pages} {565} (\bibinfo {year} {2001})}\BibitemShut {NoStop}%
\bibitem [{\citenamefont {Walther}\ \emph {et~al.}(2006)\citenamefont
  {Walther}, \citenamefont {Varcoe}, \citenamefont {Englert},\ and\
  \citenamefont {Becker}}]{Walther2006}%
  \BibitemOpen
  \bibfield  {author} {\bibinfo {author} {\bibfnamefont {H.}~\bibnamefont
  {Walther}}, \bibinfo {author} {\bibfnamefont {B.~T.~H.}\ \bibnamefont
  {Varcoe}}, \bibinfo {author} {\bibfnamefont {B.-G.}\ \bibnamefont {Englert}},
  \ and\ \bibinfo {author} {\bibfnamefont {T.}~\bibnamefont {Becker}},\ }\href
  {\doibase 10.1088/0034-4885/69/5/R02} {\bibfield  {journal} {\bibinfo
  {journal} {Reports on Progress in Physics}\ }\textbf {\bibinfo {volume}
  {69}},\ \bibinfo {pages} {1325} (\bibinfo {year} {2006})}\BibitemShut
  {NoStop}%
\bibitem [{\citenamefont {Milburn}(1989)}]{Milburn1989}%
  \BibitemOpen
  \bibfield  {author} {\bibinfo {author} {\bibfnamefont {G.~J.}\ \bibnamefont
  {Milburn}},\ }\href {\doibase 10.1103/PhysRevLett.62.2124} {\bibfield
  {journal} {\bibinfo  {journal} {Physical Review Letters}\ }\textbf {\bibinfo
  {volume} {62}},\ \bibinfo {pages} {2124} (\bibinfo {year}
  {1989})}\BibitemShut {NoStop}%
\bibitem [{\citenamefont {Yavuz}(2005)}]{Yavuz2005}%
  \BibitemOpen
  \bibfield  {author} {\bibinfo {author} {\bibfnamefont {D.~D.}\ \bibnamefont
  {Yavuz}},\ }\href {\doibase 10.1103/PhysRevA.71.053816} {\bibfield  {journal}
  {\bibinfo  {journal} {Physical Review A}\ }\textbf {\bibinfo {volume} {71}},\
  \bibinfo {pages} {053816} (\bibinfo {year} {2005})}\BibitemShut {NoStop}%
\bibitem [{\citenamefont {O'Brien}(2007)}]{OBrien2007}%
  \BibitemOpen
  \bibfield  {author} {\bibinfo {author} {\bibfnamefont {J.~L.}\ \bibnamefont
  {O'Brien}},\ }\href {\doibase 10.1126/science.1142892} {\bibfield  {journal}
  {\bibinfo  {journal} {Science}\ }\textbf {\bibinfo {volume} {318}},\ \bibinfo
  {pages} {1567} (\bibinfo {year} {2007})}\BibitemShut {NoStop}%
\bibitem [{\citenamefont {Knill}\ \emph {et~al.}(2001)\citenamefont {Knill},
  \citenamefont {Laflamme},\ and\ \citenamefont {Milburn}}]{Knill2001}%
  \BibitemOpen
  \bibfield  {author} {\bibinfo {author} {\bibfnamefont {E.}~\bibnamefont
  {Knill}}, \bibinfo {author} {\bibfnamefont {R.}~\bibnamefont {Laflamme}}, \
  and\ \bibinfo {author} {\bibfnamefont {G.~J.}\ \bibnamefont {Milburn}},\
  }\href {\doibase 10.1038/35051009} {\bibfield  {journal} {\bibinfo  {journal}
  {Nature}\ }\textbf {\bibinfo {volume} {409}},\ \bibinfo {pages} {46}
  (\bibinfo {year} {2001})}\BibitemShut {NoStop}%
\bibitem [{\citenamefont {Kok}\ \emph {et~al.}(2007)\citenamefont {Kok},
  \citenamefont {Munro}, \citenamefont {Nemoto}, \citenamefont {Ralph},
  \citenamefont {Dowling},\ and\ \citenamefont {Milburn}}]{Kok2007}%
  \BibitemOpen
  \bibfield  {author} {\bibinfo {author} {\bibfnamefont {P.}~\bibnamefont
  {Kok}}, \bibinfo {author} {\bibfnamefont {W.}~\bibnamefont {Munro}}, \bibinfo
  {author} {\bibfnamefont {K.}~\bibnamefont {Nemoto}}, \bibinfo {author}
  {\bibfnamefont {T.}~\bibnamefont {Ralph}}, \bibinfo {author} {\bibfnamefont
  {J.}~\bibnamefont {Dowling}}, \ and\ \bibinfo {author} {\bibfnamefont
  {G.~J.}\ \bibnamefont {Milburn}},\ }\href {\doibase
  10.1103/RevModPhys.79.135} {\bibfield  {journal} {\bibinfo  {journal}
  {Reviews of Modern Physics}\ }\textbf {\bibinfo {volume} {79}},\ \bibinfo
  {pages} {135} (\bibinfo {year} {2007})}\BibitemShut {NoStop}%
\bibitem [{\citenamefont {Duan}\ and\ \citenamefont {Kimble}(2004)}]{Duan2004}%
  \BibitemOpen
  \bibfield  {author} {\bibinfo {author} {\bibfnamefont {L.-M.}\ \bibnamefont
  {Duan}}\ and\ \bibinfo {author} {\bibfnamefont {H.~J.}\ \bibnamefont
  {Kimble}},\ }\href {\doibase 10.1103/PhysRevLett.92.127902} {\bibfield
  {journal} {\bibinfo  {journal} {Physical Review Letters}\ }\textbf {\bibinfo
  {volume} {92}},\ \bibinfo {pages} {127902} (\bibinfo {year}
  {2004})}\BibitemShut {NoStop}%
\bibitem [{\citenamefont {Pellizzari}\ \emph {et~al.}(1995)\citenamefont
  {Pellizzari}, \citenamefont {Gardiner}, \citenamefont {Cirac},\ and\
  \citenamefont {Zoller}}]{Pellizzari1995}%
  \BibitemOpen
  \bibfield  {author} {\bibinfo {author} {\bibfnamefont {T.}~\bibnamefont
  {Pellizzari}}, \bibinfo {author} {\bibfnamefont {S.~A.}\ \bibnamefont
  {Gardiner}}, \bibinfo {author} {\bibfnamefont {J.~I.}\ \bibnamefont {Cirac}},
  \ and\ \bibinfo {author} {\bibfnamefont {P.}~\bibnamefont {Zoller}},\ }\href
  {\doibase 10.1103/PhysRevLett.75.3788} {\bibfield  {journal} {\bibinfo
  {journal} {Physical Review Letters}\ }\textbf {\bibinfo {volume} {75}},\
  \bibinfo {pages} {3788} (\bibinfo {year} {1995})}\BibitemShut {NoStop}%
\bibitem [{\citenamefont {Rauschenbeutel}\ \emph {et~al.}(1999)\citenamefont
  {Rauschenbeutel}, \citenamefont {Nogues}, \citenamefont {Osnaghi},
  \citenamefont {Bertet}, \citenamefont {Brune}, \citenamefont {Raimond},\ and\
  \citenamefont {Haroche}}]{Rauschenbeutel1999}%
  \BibitemOpen
  \bibfield  {author} {\bibinfo {author} {\bibfnamefont {A.}~\bibnamefont
  {Rauschenbeutel}}, \bibinfo {author} {\bibfnamefont {G.}~\bibnamefont
  {Nogues}}, \bibinfo {author} {\bibfnamefont {S.}~\bibnamefont {Osnaghi}},
  \bibinfo {author} {\bibfnamefont {P.}~\bibnamefont {Bertet}}, \bibinfo
  {author} {\bibfnamefont {M.}~\bibnamefont {Brune}}, \bibinfo {author}
  {\bibfnamefont {J.~M.}\ \bibnamefont {Raimond}}, \ and\ \bibinfo {author}
  {\bibfnamefont {S.}~\bibnamefont {Haroche}},\ }\href {\doibase
  10.1103/PhysRevLett.83.5166} {\bibfield  {journal} {\bibinfo  {journal}
  {Physical Review Letters}\ }\textbf {\bibinfo {volume} {83}},\ \bibinfo
  {pages} {5166} (\bibinfo {year} {1999})}\BibitemShut {NoStop}%
\bibitem [{\citenamefont {Lin}\ \emph {et~al.}(2006)\citenamefont {Lin},
  \citenamefont {Zhou}, \citenamefont {Ye}, \citenamefont {Xiao},\ and\
  \citenamefont {Guo}}]{Lin2006}%
  \BibitemOpen
  \bibfield  {author} {\bibinfo {author} {\bibfnamefont {X.-M.}\ \bibnamefont
  {Lin}}, \bibinfo {author} {\bibfnamefont {Z.-W.}\ \bibnamefont {Zhou}},
  \bibinfo {author} {\bibfnamefont {M.-Y.}\ \bibnamefont {Ye}}, \bibinfo
  {author} {\bibfnamefont {Y.-F.}\ \bibnamefont {Xiao}}, \ and\ \bibinfo
  {author} {\bibfnamefont {G.-C.}\ \bibnamefont {Guo}},\ }\href {\doibase
  10.1103/PhysRevA.73.012323} {\bibfield  {journal} {\bibinfo  {journal}
  {Physical Review A}\ }\textbf {\bibinfo {volume} {73}},\ \bibinfo {pages}
  {012323} (\bibinfo {year} {2006})}\BibitemShut {NoStop}%
\bibitem [{\citenamefont {Biswas}\ and\ \citenamefont
  {Agarwal}(2004)}]{Biswas2004}%
  \BibitemOpen
  \bibfield  {author} {\bibinfo {author} {\bibfnamefont {A.}~\bibnamefont
  {Biswas}}\ and\ \bibinfo {author} {\bibfnamefont {G.~S.}\ \bibnamefont
  {Agarwal}},\ }\href {\doibase 10.1103/PhysRevA.69.062306} {\bibfield
  {journal} {\bibinfo  {journal} {Physical Review A}\ }\textbf {\bibinfo
  {volume} {69}},\ \bibinfo {pages} {062306} (\bibinfo {year}
  {2004})}\BibitemShut {NoStop}%
\bibitem [{\citenamefont {Joshi}\ and\ \citenamefont {Xiao}(2006)}]{Joshi2006}%
  \BibitemOpen
  \bibfield  {author} {\bibinfo {author} {\bibfnamefont {A.}~\bibnamefont
  {Joshi}}\ and\ \bibinfo {author} {\bibfnamefont {M.}~\bibnamefont {Xiao}},\
  }\href {\doibase 10.1103/PhysRevA.74.052318} {\bibfield  {journal} {\bibinfo
  {journal} {Physical Review A}\ }\textbf {\bibinfo {volume} {74}},\ \bibinfo
  {pages} {052318} (\bibinfo {year} {2006})}\BibitemShut {NoStop}%
\bibitem [{\citenamefont {Zubairy}\ \emph {et~al.}(2003)\citenamefont
  {Zubairy}, \citenamefont {Kim},\ and\ \citenamefont {Scully}}]{Zubairy2003}%
  \BibitemOpen
  \bibfield  {author} {\bibinfo {author} {\bibfnamefont {M.~S.}\ \bibnamefont
  {Zubairy}}, \bibinfo {author} {\bibfnamefont {M.}~\bibnamefont {Kim}}, \ and\
  \bibinfo {author} {\bibfnamefont {M.~O.}\ \bibnamefont {Scully}},\ }\href
  {\doibase 10.1103/PhysRevA.68.033820} {\bibfield  {journal} {\bibinfo
  {journal} {Physical Review A}\ }\textbf {\bibinfo {volume} {68}},\ \bibinfo
  {pages} {033820} (\bibinfo {year} {2003})}\BibitemShut {NoStop}%
\bibitem [{\citenamefont {Shu}\ \emph {et~al.}(2007)\citenamefont {Shu},
  \citenamefont {Zou}, \citenamefont {Xiao},\ and\ \citenamefont
  {Guo}}]{Shu2007}%
  \BibitemOpen
  \bibfield  {author} {\bibinfo {author} {\bibfnamefont {J.}~\bibnamefont
  {Shu}}, \bibinfo {author} {\bibfnamefont {X.-B.}\ \bibnamefont {Zou}},
  \bibinfo {author} {\bibfnamefont {Y.-F.}\ \bibnamefont {Xiao}}, \ and\
  \bibinfo {author} {\bibfnamefont {G.-C.}\ \bibnamefont {Guo}},\ }\href
  {\doibase 10.1103/PhysRevA.75.044302} {\bibfield  {journal} {\bibinfo
  {journal} {Physical Review A}\ }\textbf {\bibinfo {volume} {75}},\ \bibinfo
  {pages} {044302} (\bibinfo {year} {2007})}\BibitemShut {NoStop}%
\bibitem [{\citenamefont {Lin}\ \emph {et~al.}(2008)\citenamefont {Lin},
  \citenamefont {Zou}, \citenamefont {Ye}, \citenamefont {Lin},\ and\
  \citenamefont {Guo}}]{Lin2008}%
  \BibitemOpen
  \bibfield  {author} {\bibinfo {author} {\bibfnamefont {G.-W.}\ \bibnamefont
  {Lin}}, \bibinfo {author} {\bibfnamefont {X.-B.}\ \bibnamefont {Zou}},
  \bibinfo {author} {\bibfnamefont {M.-Y.}\ \bibnamefont {Ye}}, \bibinfo
  {author} {\bibfnamefont {X.-M.}\ \bibnamefont {Lin}}, \ and\ \bibinfo
  {author} {\bibfnamefont {G.-C.}\ \bibnamefont {Guo}},\ }\href {\doibase
  10.1103/PhysRevA.77.064301} {\bibfield  {journal} {\bibinfo  {journal}
  {Physical Review A}\ }\textbf {\bibinfo {volume} {77}},\ \bibinfo {pages}
  {064301} (\bibinfo {year} {2008})}\BibitemShut {NoStop}%
\bibitem [{\citenamefont {Chuang}\ and\ \citenamefont
  {Yamamoto}(1995)}]{Chuang1995}%
  \BibitemOpen
  \bibfield  {author} {\bibinfo {author} {\bibfnamefont {I.~L.}\ \bibnamefont
  {Chuang}}\ and\ \bibinfo {author} {\bibfnamefont {Y.}~\bibnamefont
  {Yamamoto}},\ }\href {\doibase 10.1103/PhysRevA.52.3489} {\bibfield
  {journal} {\bibinfo  {journal} {Physical Review A}\ }\textbf {\bibinfo
  {volume} {52}},\ \bibinfo {pages} {3489} (\bibinfo {year}
  {1995})}\BibitemShut {NoStop}%
\bibitem [{\citenamefont {Bellomo}\ \emph {et~al.}(2007)\citenamefont
  {Bellomo}, \citenamefont {Lo~Franco},\ and\ \citenamefont
  {Compagno}}]{Bellomo07}%
  \BibitemOpen
  \bibfield  {author} {\bibinfo {author} {\bibfnamefont {B.}~\bibnamefont
  {Bellomo}}, \bibinfo {author} {\bibfnamefont {R.}~\bibnamefont {Lo~Franco}},
  \ and\ \bibinfo {author} {\bibfnamefont {G.}~\bibnamefont {Compagno}},\
  }\href {\doibase 10.1103/PhysRevLett.99.160502} {\bibfield  {journal}
  {\bibinfo  {journal} {Physical Review Letters}\ }\textbf {\bibinfo {volume}
  {99}},\ \bibinfo {pages} {160502} (\bibinfo {year} {2007})}\BibitemShut
  {NoStop}%
\bibitem [{\citenamefont {Maniscalco}\ \emph {et~al.}(2008)\citenamefont
  {Maniscalco}, \citenamefont {Francica}, \citenamefont {Zaffino},
  \citenamefont {Lo~Gullo},\ and\ \citenamefont {Plastina}}]{Maniscalco08}%
  \BibitemOpen
  \bibfield  {author} {\bibinfo {author} {\bibfnamefont {S.}~\bibnamefont
  {Maniscalco}}, \bibinfo {author} {\bibfnamefont {F.}~\bibnamefont
  {Francica}}, \bibinfo {author} {\bibfnamefont {R.~L.}\ \bibnamefont
  {Zaffino}}, \bibinfo {author} {\bibfnamefont {N.}~\bibnamefont {Lo~Gullo}}, \
  and\ \bibinfo {author} {\bibfnamefont {F.}~\bibnamefont {Plastina}},\ }\href
  {\doibase 10.1103/PhysRevLett.100.090503} {\bibfield  {journal} {\bibinfo
  {journal} {Physical Review Letters}\ }\textbf {\bibinfo {volume} {100}},\
  \bibinfo {pages} {090503} (\bibinfo {year} {2008})}\BibitemShut {NoStop}%
\bibitem [{\citenamefont {Wang}\ \emph {et~al.}(2008)\citenamefont {Wang},
  \citenamefont {Zhang},\ and\ \citenamefont {Liang}}]{Wang08}%
  \BibitemOpen
  \bibfield  {author} {\bibinfo {author} {\bibfnamefont {F.-Q.}\ \bibnamefont
  {Wang}}, \bibinfo {author} {\bibfnamefont {Z.-M.}\ \bibnamefont {Zhang}}, \
  and\ \bibinfo {author} {\bibfnamefont {R.-S.}\ \bibnamefont {Liang}},\ }\href
  {\doibase 10.1103/PhysRevA.78.042320} {\bibfield  {journal} {\bibinfo
  {journal} {Physical Review A}\ }\textbf {\bibinfo {volume} {78}},\ \bibinfo
  {pages} {042320} (\bibinfo {year} {2008})}\BibitemShut {NoStop}%
\bibitem [{\citenamefont {Li}\ \emph {et~al.}(2009)\citenamefont {Li},
  \citenamefont {Zhou},\ and\ \citenamefont {Guo}}]{Li09}%
  \BibitemOpen
  \bibfield  {author} {\bibinfo {author} {\bibfnamefont {Y.}~\bibnamefont
  {Li}}, \bibinfo {author} {\bibfnamefont {J.}~\bibnamefont {Zhou}}, \ and\
  \bibinfo {author} {\bibfnamefont {H.}~\bibnamefont {Guo}},\ }\href {\doibase
  10.1103/PhysRevA.79.012309} {\bibfield  {journal} {\bibinfo  {journal}
  {Physical Review A}\ }\textbf {\bibinfo {volume} {79}},\ \bibinfo {pages}
  {012309} (\bibinfo {year} {2009})}\BibitemShut {NoStop}%
\bibitem [{\citenamefont {Mazzola}\ \emph
  {et~al.}(2009{\natexlab{a}})\citenamefont {Mazzola}, \citenamefont
  {Maniscalco}, \citenamefont {Piilo}, \citenamefont {Suominen},\ and\
  \citenamefont {Garraway}}]{Mazzola09a}%
  \BibitemOpen
  \bibfield  {author} {\bibinfo {author} {\bibfnamefont {L.}~\bibnamefont
  {Mazzola}}, \bibinfo {author} {\bibfnamefont {S.}~\bibnamefont {Maniscalco}},
  \bibinfo {author} {\bibfnamefont {J.}~\bibnamefont {Piilo}}, \bibinfo
  {author} {\bibfnamefont {K.-A.}\ \bibnamefont {Suominen}}, \ and\ \bibinfo
  {author} {\bibfnamefont {B.~M.}\ \bibnamefont {Garraway}},\ }\href {\doibase
  10.1103/PhysRevA.79.042302} {\bibfield  {journal} {\bibinfo  {journal}
  {Physical Review A}\ }\textbf {\bibinfo {volume} {79}},\ \bibinfo {pages}
  {042302} (\bibinfo {year} {2009}{\natexlab{a}})}\BibitemShut {NoStop}%
\bibitem [{\citenamefont {Zhou}\ \emph {et~al.}(2009)\citenamefont {Zhou},
  \citenamefont {Wu}, \citenamefont {Zhu},\ and\ \citenamefont {Guo}}]{Zhou09}%
  \BibitemOpen
  \bibfield  {author} {\bibinfo {author} {\bibfnamefont {J.}~\bibnamefont
  {Zhou}}, \bibinfo {author} {\bibfnamefont {C.}~\bibnamefont {Wu}}, \bibinfo
  {author} {\bibfnamefont {M.}~\bibnamefont {Zhu}}, \ and\ \bibinfo {author}
  {\bibfnamefont {H.}~\bibnamefont {Guo}},\ }\href {\doibase
  10.1088/0953-4075/42/21/215505} {\bibfield  {journal} {\bibinfo  {journal}
  {Journal of Physics B: Atomic, Molecular and Optical Physics}\ }\textbf
  {\bibinfo {volume} {42}},\ \bibinfo {pages} {215505} (\bibinfo {year}
  {2009})}\BibitemShut {NoStop}%
\bibitem [{\citenamefont {Ferraro}\ \emph {et~al.}(2009)\citenamefont
  {Ferraro}, \citenamefont {Scala}, \citenamefont {Migliore},\ and\
  \citenamefont {Napoli}}]{Ferraro09}%
  \BibitemOpen
  \bibfield  {author} {\bibinfo {author} {\bibfnamefont {E.}~\bibnamefont
  {Ferraro}}, \bibinfo {author} {\bibfnamefont {M.}~\bibnamefont {Scala}},
  \bibinfo {author} {\bibfnamefont {R.}~\bibnamefont {Migliore}}, \ and\
  \bibinfo {author} {\bibfnamefont {A.}~\bibnamefont {Napoli}},\ }\href
  {\doibase 10.1103/PhysRevA.80.042112} {\bibfield  {journal} {\bibinfo
  {journal} {Physical Review A}\ }\textbf {\bibinfo {volume} {80}},\ \bibinfo
  {pages} {042112} (\bibinfo {year} {2009})}\BibitemShut {NoStop}%
\bibitem [{\citenamefont {Francica}\ \emph {et~al.}(2009)\citenamefont
  {Francica}, \citenamefont {Maniscalco}, \citenamefont {Piilo}, \citenamefont
  {Plastina},\ and\ \citenamefont {Suominen}}]{Francica09}%
  \BibitemOpen
  \bibfield  {author} {\bibinfo {author} {\bibfnamefont {F.}~\bibnamefont
  {Francica}}, \bibinfo {author} {\bibfnamefont {S.}~\bibnamefont
  {Maniscalco}}, \bibinfo {author} {\bibfnamefont {J.}~\bibnamefont {Piilo}},
  \bibinfo {author} {\bibfnamefont {F.}~\bibnamefont {Plastina}}, \ and\
  \bibinfo {author} {\bibfnamefont {K.-A.}\ \bibnamefont {Suominen}},\ }\href
  {\doibase 10.1103/PhysRevA.79.032310} {\bibfield  {journal} {\bibinfo
  {journal} {Physical Review A}\ }\textbf {\bibinfo {volume} {79}},\ \bibinfo
  {pages} {032310} (\bibinfo {year} {2009})}\BibitemShut {NoStop}%
\bibitem [{\citenamefont {Xiao}\ \emph {et~al.}(2009)\citenamefont {Xiao},
  \citenamefont {Fang}, \citenamefont {Li}, \citenamefont {Zeng},\ and\
  \citenamefont {Wu}}]{Xiao09}%
  \BibitemOpen
  \bibfield  {author} {\bibinfo {author} {\bibfnamefont {X.}~\bibnamefont
  {Xiao}}, \bibinfo {author} {\bibfnamefont {M.-F.}\ \bibnamefont {Fang}},
  \bibinfo {author} {\bibfnamefont {Y.-L.}\ \bibnamefont {Li}}, \bibinfo
  {author} {\bibfnamefont {K.}~\bibnamefont {Zeng}}, \ and\ \bibinfo {author}
  {\bibfnamefont {C.}~\bibnamefont {Wu}},\ }\href {\doibase
  10.1088/0953-4075/42/23/235502} {\bibfield  {journal} {\bibinfo  {journal}
  {Journal of Physics B: Atomic, Molecular and Optical Physics}\ }\textbf
  {\bibinfo {volume} {42}},\ \bibinfo {pages} {235502} (\bibinfo {year}
  {2009})}\BibitemShut {NoStop}%
\bibitem [{\citenamefont {Zhang}\ \emph {et~al.}(2009)\citenamefont {Zhang},
  \citenamefont {Man},\ and\ \citenamefont {Xia}}]{Zhang09}%
  \BibitemOpen
  \bibfield  {author} {\bibinfo {author} {\bibfnamefont {Y.~J.}\ \bibnamefont
  {Zhang}}, \bibinfo {author} {\bibfnamefont {Z.~X.}\ \bibnamefont {Man}}, \
  and\ \bibinfo {author} {\bibfnamefont {Y.~J.}\ \bibnamefont {Xia}},\ }\href
  {\doibase 10.1140/epjd/e2009-00226-2} {\bibfield  {journal} {\bibinfo
  {journal} {The European Physical Journal D}\ }\textbf {\bibinfo {volume}
  {55}},\ \bibinfo {pages} {173} (\bibinfo {year} {2009})}\BibitemShut
  {NoStop}%
\bibitem [{\citenamefont {Mazzola}\ \emph
  {et~al.}(2009{\natexlab{b}})\citenamefont {Mazzola}, \citenamefont
  {Maniscalco}, \citenamefont {Suominen},\ and\ \citenamefont
  {Garraway}}]{Mazzola09c}%
  \BibitemOpen
  \bibfield  {author} {\bibinfo {author} {\bibfnamefont {L.}~\bibnamefont
  {Mazzola}}, \bibinfo {author} {\bibfnamefont {S.}~\bibnamefont {Maniscalco}},
  \bibinfo {author} {\bibfnamefont {K.-A.}\ \bibnamefont {Suominen}}, \ and\
  \bibinfo {author} {\bibfnamefont {B.~M.}\ \bibnamefont {Garraway}},\ }\href
  {\doibase 10.1007/s11128-009-0135-8} {\bibfield  {journal} {\bibinfo
  {journal} {Quantum Information Processing}\ }\textbf {\bibinfo {volume}
  {8}},\ \bibinfo {pages} {577} (\bibinfo {year}
  {2009}{\natexlab{b}})}\BibitemShut {NoStop}%
\bibitem [{\citenamefont {Mazzola}\ \emph
  {et~al.}(2010{\natexlab{a}})\citenamefont {Mazzola}, \citenamefont
  {Maniscalco}, \citenamefont {Piilo},\ and\ \citenamefont
  {Suominen}}]{Mazzola10b}%
  \BibitemOpen
  \bibfield  {author} {\bibinfo {author} {\bibfnamefont {L.}~\bibnamefont
  {Mazzola}}, \bibinfo {author} {\bibfnamefont {S.}~\bibnamefont {Maniscalco}},
  \bibinfo {author} {\bibfnamefont {J.}~\bibnamefont {Piilo}}, \ and\ \bibinfo
  {author} {\bibfnamefont {K.-A.}\ \bibnamefont {Suominen}},\ }\href {\doibase
  10.1088/0953-4075/43/8/085505} {\bibfield  {journal} {\bibinfo  {journal}
  {Journal of Physics B: Atomic, Molecular and Optical Physics}\ }\textbf
  {\bibinfo {volume} {43}},\ \bibinfo {pages} {085505} (\bibinfo {year}
  {2010}{\natexlab{a}})}\BibitemShut {NoStop}%
\bibitem [{\citenamefont {Mazzola}\ \emph
  {et~al.}(2010{\natexlab{b}})\citenamefont {Mazzola}, \citenamefont {Bellomo},
  \citenamefont {Lo~Franco},\ and\ \citenamefont {Compagno}}]{Mazzola10a}%
  \BibitemOpen
  \bibfield  {author} {\bibinfo {author} {\bibfnamefont {L.}~\bibnamefont
  {Mazzola}}, \bibinfo {author} {\bibfnamefont {B.}~\bibnamefont {Bellomo}},
  \bibinfo {author} {\bibfnamefont {R.}~\bibnamefont {Lo~Franco}}, \ and\
  \bibinfo {author} {\bibfnamefont {G.}~\bibnamefont {Compagno}},\ }\href
  {\doibase 10.1103/PhysRevA.81.052116} {\bibfield  {journal} {\bibinfo
  {journal} {Physical Review A}\ }\textbf {\bibinfo {volume} {81}},\ \bibinfo
  {pages} {052116} (\bibinfo {year} {2010}{\natexlab{b}})}\BibitemShut
  {NoStop}%
\bibitem [{\citenamefont {Fanchini}\ \emph {et~al.}(2010)\citenamefont
  {Fanchini}, \citenamefont {Werlang}, \citenamefont {Brasil}, \citenamefont
  {Arruda},\ and\ \citenamefont {Caldeira}}]{Fanchini10}%
  \BibitemOpen
  \bibfield  {author} {\bibinfo {author} {\bibfnamefont {F.~F.}\ \bibnamefont
  {Fanchini}}, \bibinfo {author} {\bibfnamefont {T.}~\bibnamefont {Werlang}},
  \bibinfo {author} {\bibfnamefont {C.~A.}\ \bibnamefont {Brasil}}, \bibinfo
  {author} {\bibfnamefont {L.~G.~E.}\ \bibnamefont {Arruda}}, \ and\ \bibinfo
  {author} {\bibfnamefont {A.~O.}\ \bibnamefont {Caldeira}},\ }\href {\doibase
  10.1103/PhysRevA.81.052107} {\bibfield  {journal} {\bibinfo  {journal}
  {Physical Review A}\ }\textbf {\bibinfo {volume} {81}},\ \bibinfo {pages}
  {052107} (\bibinfo {year} {2010})}\BibitemShut {NoStop}%
\bibitem [{\citenamefont {Tong}\ \emph {et~al.}(2010)\citenamefont {Tong},
  \citenamefont {An}, \citenamefont {Luo},\ and\ \citenamefont {Oh}}]{Ge10}%
  \BibitemOpen
  \bibfield  {author} {\bibinfo {author} {\bibfnamefont {Q.-J.}\ \bibnamefont
  {Tong}}, \bibinfo {author} {\bibfnamefont {J.-H.}\ \bibnamefont {An}},
  \bibinfo {author} {\bibfnamefont {H.-G.}\ \bibnamefont {Luo}}, \ and\
  \bibinfo {author} {\bibfnamefont {C.~H.}\ \bibnamefont {Oh}},\ }\href
  {\doibase 10.1103/PhysRevA.81.052330} {\bibfield  {journal} {\bibinfo
  {journal} {Physical Review A}\ }\textbf {\bibinfo {volume} {81}},\ \bibinfo
  {pages} {052330} (\bibinfo {year} {2010})}\BibitemShut {NoStop}%
\bibitem [{\citenamefont {Wang}\ \emph {et~al.}(2010)\citenamefont {Wang},
  \citenamefont {Xu}, \citenamefont {Chen},\ and\ \citenamefont
  {Feng}}]{Wang10}%
  \BibitemOpen
  \bibfield  {author} {\bibinfo {author} {\bibfnamefont {B.}~\bibnamefont
  {Wang}}, \bibinfo {author} {\bibfnamefont {Z.-Y.}\ \bibnamefont {Xu}},
  \bibinfo {author} {\bibfnamefont {Z.-Q.}\ \bibnamefont {Chen}}, \ and\
  \bibinfo {author} {\bibfnamefont {M.}~\bibnamefont {Feng}},\ }\href {\doibase
  10.1103/PhysRevA.81.014101} {\bibfield  {journal} {\bibinfo  {journal}
  {Physical Review A}\ }\textbf {\bibinfo {volume} {81}},\ \bibinfo {pages}
  {014101} (\bibinfo {year} {2010})}\BibitemShut {NoStop}%
\bibitem [{\citenamefont {Man}\ \emph {et~al.}(2010)\citenamefont {Man},
  \citenamefont {Zhang}, \citenamefont {Su},\ and\ \citenamefont
  {Xia}}]{Man10}%
  \BibitemOpen
  \bibfield  {author} {\bibinfo {author} {\bibfnamefont {Z.~X.}\ \bibnamefont
  {Man}}, \bibinfo {author} {\bibfnamefont {Y.~J.}\ \bibnamefont {Zhang}},
  \bibinfo {author} {\bibfnamefont {F.}~\bibnamefont {Su}}, \ and\ \bibinfo
  {author} {\bibfnamefont {Y.~J.}\ \bibnamefont {Xia}},\ }\href {\doibase
  10.1140/epjd/e2010-00094-7} {\bibfield  {journal} {\bibinfo  {journal} {The
  European Physical Journal D}\ }\textbf {\bibinfo {volume} {58}},\ \bibinfo
  {pages} {147} (\bibinfo {year} {2010})}\BibitemShut {NoStop}%
\bibitem [{\citenamefont {Meschede}\ \emph {et~al.}(1985)\citenamefont
  {Meschede}, \citenamefont {Walther},\ and\ \citenamefont
  {M\"{u}ller}}]{Meschede1985}%
  \BibitemOpen
  \bibfield  {author} {\bibinfo {author} {\bibfnamefont {D.}~\bibnamefont
  {Meschede}}, \bibinfo {author} {\bibfnamefont {H.}~\bibnamefont {Walther}}, \
  and\ \bibinfo {author} {\bibfnamefont {G.}~\bibnamefont {M\"{u}ller}},\
  }\href {\doibase 10.1103/PhysRevLett.54.551} {\bibfield  {journal} {\bibinfo
  {journal} {Physical Review Letters}\ }\textbf {\bibinfo {volume} {54}},\
  \bibinfo {pages} {551} (\bibinfo {year} {1985})}\BibitemShut {NoStop}%
\bibitem [{\citenamefont {Zheng}\ and\ \citenamefont {Guo}(2000)}]{Zheng2000}%
  \BibitemOpen
  \bibfield  {author} {\bibinfo {author} {\bibfnamefont {S.-B.}\ \bibnamefont
  {Zheng}}\ and\ \bibinfo {author} {\bibfnamefont {G.-C.}\ \bibnamefont
  {Guo}},\ }\href {\doibase 10.1103/PhysRevLett.85.2392} {\bibfield  {journal}
  {\bibinfo  {journal} {Physical Review Letters}\ }\textbf {\bibinfo {volume}
  {85}},\ \bibinfo {pages} {2392} (\bibinfo {year} {2000})}\BibitemShut
  {NoStop}%
\bibitem [{\citenamefont {Osnaghi}\ \emph {et~al.}(2001)\citenamefont
  {Osnaghi}, \citenamefont {Bertet}, \citenamefont {Auffeves}, \citenamefont
  {Maioli}, \citenamefont {Brune}, \citenamefont {Raimond},\ and\ \citenamefont
  {Haroche}}]{Osnaghi2001}%
  \BibitemOpen
  \bibfield  {author} {\bibinfo {author} {\bibfnamefont {S.}~\bibnamefont
  {Osnaghi}}, \bibinfo {author} {\bibfnamefont {P.}~\bibnamefont {Bertet}},
  \bibinfo {author} {\bibfnamefont {A.}~\bibnamefont {Auffeves}}, \bibinfo
  {author} {\bibfnamefont {P.}~\bibnamefont {Maioli}}, \bibinfo {author}
  {\bibfnamefont {M.}~\bibnamefont {Brune}}, \bibinfo {author} {\bibfnamefont
  {J.~M.}\ \bibnamefont {Raimond}}, \ and\ \bibinfo {author} {\bibfnamefont
  {S.}~\bibnamefont {Haroche}},\ }\href {\doibase
  10.1103/PhysRevLett.87.037902} {\bibfield  {journal} {\bibinfo  {journal}
  {Physical Review Letters}\ }\textbf {\bibinfo {volume} {87}},\ \bibinfo
  {pages} {037902} (\bibinfo {year} {2001})}\BibitemShut {NoStop}%
\bibitem [{\citenamefont {Garraway}\ \emph {et~al.}(1994)\citenamefont
  {Garraway}, \citenamefont {Sherman}, \citenamefont {Moya-Cessa},
  \citenamefont {Knight},\ and\ \citenamefont {Kurizki}}]{Garraway1994}%
  \BibitemOpen
  \bibfield  {author} {\bibinfo {author} {\bibfnamefont {B.~M.}\ \bibnamefont
  {Garraway}}, \bibinfo {author} {\bibfnamefont {B.}~\bibnamefont {Sherman}},
  \bibinfo {author} {\bibfnamefont {H.}~\bibnamefont {Moya-Cessa}}, \bibinfo
  {author} {\bibfnamefont {P.~L.}\ \bibnamefont {Knight}}, \ and\ \bibinfo
  {author} {\bibfnamefont {G.}~\bibnamefont {Kurizki}},\ }\href {\doibase
  10.1103/PhysRevA.49.535} {\bibfield  {journal} {\bibinfo  {journal} {Physical
  Review A}\ }\textbf {\bibinfo {volume} {49}},\ \bibinfo {pages} {535}
  (\bibinfo {year} {1994})}\BibitemShut {NoStop}%
\bibitem [{\citenamefont {Franson}\ \emph {et~al.}(2004)\citenamefont
  {Franson}, \citenamefont {Jacobs},\ and\ \citenamefont
  {Pittman}}]{Franson2004}%
  \BibitemOpen
  \bibfield  {author} {\bibinfo {author} {\bibfnamefont {J.~D.}\ \bibnamefont
  {Franson}}, \bibinfo {author} {\bibfnamefont {B.~C.}\ \bibnamefont {Jacobs}},
  \ and\ \bibinfo {author} {\bibfnamefont {T.~B.}\ \bibnamefont {Pittman}},\
  }\href {\doibase 10.1103/PhysRevA.70.062302} {\bibfield  {journal} {\bibinfo
  {journal} {Physical Review A}\ }\textbf {\bibinfo {volume} {70}},\ \bibinfo
  {pages} {062302} (\bibinfo {year} {2004})}\BibitemShut {NoStop}%
\bibitem [{\citenamefont {Shore}(1981)}]{Shore1981}%
  \BibitemOpen
  \bibfield  {author} {\bibinfo {author} {\bibfnamefont {B.~W.}\ \bibnamefont
  {Shore}},\ }\href {\doibase 10.1103/PhysRevA.24.1413} {\bibfield  {journal}
  {\bibinfo  {journal} {Physical Review A}\ }\textbf {\bibinfo {volume} {24}},\
  \bibinfo {pages} {1413} (\bibinfo {year} {1981})}\BibitemShut {NoStop}%
\bibitem [{\citenamefont {Schuch}\ and\ \citenamefont
  {Siewert}(2003)}]{Schuch2003}%
  \BibitemOpen
  \bibfield  {author} {\bibinfo {author} {\bibfnamefont {N.}~\bibnamefont
  {Schuch}}\ and\ \bibinfo {author} {\bibfnamefont {J.}~\bibnamefont
  {Siewert}},\ }\href {\doibase 10.1103/PhysRevA.67.032301} {\bibfield
  {journal} {\bibinfo  {journal} {Physical Review A}\ }\textbf {\bibinfo
  {volume} {67}},\ \bibinfo {pages} {032301} (\bibinfo {year}
  {2003})}\BibitemShut {NoStop}%
\bibitem [{\citenamefont {Everitt}\ and\ \citenamefont
  {Garraway}(2007)}]{Everitt2007}%
  \BibitemOpen
  \bibfield  {author} {\bibinfo {author} {\bibfnamefont {M.~S.}\ \bibnamefont
  {Everitt}}\ and\ \bibinfo {author} {\bibfnamefont {B.~M.}\ \bibnamefont
  {Garraway}},\ }in\ \href
  {http://www.opticsinfobase.org/abstract.cfm?id=139750} {\emph {\bibinfo
  {booktitle} {{Conference on Coherence and Quantum Optics}}}},\ \bibinfo
  {series and number} {OSA Technical Digest (CD)}\ (\bibinfo  {publisher}
  {Optical Society of America},\ \bibinfo {year} {2007})\ pp.\ \bibinfo {pages}
  {456--457}\BibitemShut {NoStop}%
\bibitem [{\citenamefont {Cook}\ and\ \citenamefont {Shore}(1979)}]{Cook1979}%
  \BibitemOpen
  \bibfield  {author} {\bibinfo {author} {\bibfnamefont {R.~J.}\ \bibnamefont
  {Cook}}\ and\ \bibinfo {author} {\bibfnamefont {B.~W.}\ \bibnamefont
  {Shore}},\ }\href {\doibase 10.1103/PhysRevA.20.539} {\bibfield  {journal}
  {\bibinfo  {journal} {Physical Review A}\ }\textbf {\bibinfo {volume} {20}},\
  \bibinfo {pages} {539} (\bibinfo {year} {1979})}\BibitemShut {NoStop}%
\bibitem [{\citenamefont {Englert}\ \emph {et~al.}(1998)\citenamefont
  {Englert}, \citenamefont {L\"{o}ffler}, \citenamefont {Benson}, \citenamefont
  {Varcoe}, \citenamefont {Weidinger},\ and\ \citenamefont
  {Walther}}]{Englert1998}%
  \BibitemOpen
  \bibfield  {author} {\bibinfo {author} {\bibfnamefont {B.-G.}\ \bibnamefont
  {Englert}}, \bibinfo {author} {\bibfnamefont {M.}~\bibnamefont
  {L\"{o}ffler}}, \bibinfo {author} {\bibfnamefont {O.}~\bibnamefont {Benson}},
  \bibinfo {author} {\bibfnamefont {B.}~\bibnamefont {Varcoe}}, \bibinfo
  {author} {\bibfnamefont {M.}~\bibnamefont {Weidinger}}, \ and\ \bibinfo
  {author} {\bibfnamefont {H.}~\bibnamefont {Walther}},\ }\href {\doibase
  10.1002/(SICI)1521-3978(199811)46:6/8\%3C897::AID-PROP897\%3E3.0.CO;2-2}
  {\bibfield  {journal} {\bibinfo  {journal} {Fortschritte der Physik}\
  }\textbf {\bibinfo {volume} {46}},\ \bibinfo {pages} {897} (\bibinfo {year}
  {1998})}\BibitemShut {NoStop}%
\bibitem [{Note1()}]{Note1}%
  \BibitemOpen
  \bibinfo {note} {This may be a dipole forbidden transition, or an allowed
  two-photon transition, for example.}\BibitemShut {Stop}%
\bibitem [{\citenamefont {Alqahtani}\ \emph {et~al.}(2014)\citenamefont
  {Alqahtani}, \citenamefont {Everitt},\ and\ \citenamefont
  {Garraway}}]{Everitt14}%
  \BibitemOpen
  \bibfield  {author} {\bibinfo {author} {\bibfnamefont {M.~M.}\ \bibnamefont
  {Alqahtani}}, \bibinfo {author} {\bibfnamefont {M.~S.}\ \bibnamefont
  {Everitt}}, \ and\ \bibinfo {author} {\bibfnamefont {B.~M.}\ \bibnamefont
  {Garraway}},\ }\href@noop {} {\bibfield  {journal} {\bibinfo  {journal}
  {submitted}\ } (\bibinfo {year} {2014})}\BibitemShut {NoStop}%
\bibitem [{\citenamefont {Larson}\ and\ \citenamefont
  {Garraway}(2004)}]{Larson2004}%
  \BibitemOpen
  \bibfield  {author} {\bibinfo {author} {\bibfnamefont {J.}~\bibnamefont
  {Larson}}\ and\ \bibinfo {author} {\bibfnamefont {B.~M.}\ \bibnamefont
  {Garraway}},\ }\href {\doibase 10.1080/09500340408232483} {\bibfield
  {journal} {\bibinfo  {journal} {Journal of Modern Optics}\ }\textbf {\bibinfo
  {volume} {51}},\ \bibinfo {pages} {1691} (\bibinfo {year}
  {2004})}\BibitemShut {NoStop}%
\end{thebibliography}%

\end{document}